\documentclass[11pt]{article}
\usepackage{graphicx}

\title{{\Large Low-lying Dirac eigenmodes and monopoles in 3+1D compact QED}}
\author{Toru T. Takahashi
\thanks{Yukawa Institute for Theoretical Physics, Kyoto university,
Kitashirakawa-Oiwakecho, Sakyo, Kyoto 606-8502, Japan}}

\begin{document}

\maketitle

\abstract{
We study the properties of low-lying Dirac modes
in quenched compact QED at $\beta =1.01$,
employing $12^3\times N_t$ ($N_t =4,6,8,10,12$) lattices
and the overlap formalism for the fermion action.
We pay attention to the spatial distributions
of low-lying Dirac modes
below and above the ``phase transition temperature'' $T_c$.
Near-zero modes are found to have universal anti-correlations
with monopole currents, and are found to lose
their temporal structures above $T_c$
exhibiting stronger spatial localization properties.
We also study the nearest-neighbor level spacing distribution
of Dirac eigenvalues and find a Wigner-Poisson transition.
}

\section{Introduction}
\label{introduction}

In the previous paper~\cite{Takahashi:2007dv},
the spatial correlations between monopoles~\cite{DeGrand:1980eq}
and low-lying Dirac eigenmodes (eigenfunctions of the Dirac operator)
were studied in 4D quenched compact QED,
and it was found that there exist universal anti-correlations
between them below and above the critical coupling $\beta_c$.
The clear anti-correlation between Dirac eigenfunctions and monopoles 
implies that the level dynamics of Dirac eigenvalues,
which is responsible for the chiral phase transition
via Banks-Casher relation~\cite{Banks:1979yr},
is controlled by monopole configurations in 4D Euclidean 
system~\cite{Takahashi:2007dv}.
In the strong coupling phase ($\beta < \beta_c$), 
monopoles form global and complicated clusters
and make the vacuum complex
bringing about repulsive forces among Dirac 
eigenvalues~\cite{Takahashi:2007dv,Berry:1977wk,Bohigas:1983er}.
This repulsive force among eigenvalues,
which is observed as the Wigner distribution in the neighboring level spacings,
forms the non-vanishing spectral density at the spectral
origin, which is equivalent to the non-vanishing chiral condensate
$\langle\overline\psi \psi \rangle$.
On the other hand, in the weak coupling phase ($\beta > \beta_c$),
large monopole clusters vanish and the vacuum structure is much simpler,
which leads to weaker repulsive forces among 
eigenvalues~\cite{Takahashi:2007dv,Berry:1977wk,Bohigas:1983er}
and results in the Poisson statistics in the level spacings.
This weaker repulsive force is not so strong that
Dirac eigenvalues can form non-zero spectral density at the origin,
and the chiral condensate vanishes in this phase.

We in Ref.~\cite{Takahashi:2007dv} 
varied the couplings around the critical coupling $\beta_c$
employing isotropic 4D systems with the total volumes fixed,
and investigated the natures of low-lying Dirac modes.
In this case, the chirally symmetric vacuum 
realized at $\beta > \beta_c$ is rather simple and perturbative.
It is however known that the ``finite temperature'' transition
still exhibits several nonperturbative features.
For example, in the case of QCD, the formation of 
the strongly-coupled quark-gluon plasma phase 
(sQGP)~\cite{Shuryak:2003xe,Umeda:2002vr,Asakawa:2003re}
just above the transition temperature
was recently suggested and attracting many interests.
Nonperturbative aspects above the transition temperature $T_c$
can be also observed in the Wilson loops.
Whereas temporal Wilson loops exhibit deconfinement feature above $T_c$,
spatial Wilson loops still show an area law
({\it i.e.} temporally deconfined and spatially confined),
which hints that the vacuum structure is not simple but still complicated
even above $T_c$.
The (3+1)D compact QED has similar properties, and
the remaining nonperturbative structures can be also found in it.
We investigate the properties of Dirac eigenmodes,
eigenvalues, and monopoles in quenched (3+1)D compact QED,
and see what happens.

The organization of this paper is as follows:
In Sec.~\ref{formalism}, we briefly show our formalism.
Several properties of low-lying Dirac modes and eigenvalues
are clarified in Sec.~\ref{diracmodes}.
Sec.~\ref{discussions} is devoted to the discussion
based on numerical results.
We summarize the paper in Sec.~\ref{summary}.

\section{Formalism}
\label{formalism}

We adopt the Wilson gauge action at $\beta$=1.01 for gauge fields,
\begin{equation}
S_{\rm QED}
=
\beta \sum_{x}\sum_{\mu , \nu}(1-\cos\theta_{\mu\nu}(x)),
\end{equation}
and employ the overlap-Dirac operator~\cite{Neuberger:1997fp,Neuberger:1998wv},
which is constructed as
\begin{equation}
D\equiv\rho[1+\gamma_5 {\rm sgn}(H_W)]
\equiv\rho\left[1+\gamma_5 \frac{H_W}{\sqrt{{H_W}^2}}\right],
\end{equation}
and realizes the exact chiral symmetry on a lattice~\cite{Luscher:1998pq,Ginsparg:1981bj}.
Here, $H_W\equiv \gamma_5 (D_W-\rho)$  is the hermitian Wilson-Dirac operator
defined with the standard Wilson-Dirac operator $D_W$.
The ``negative mass'' $\rho$ is chosen in the range of $0<\rho <2$,
which we set 1.6 throughout this paper.
We impose
the periodic boundary conditions in all the spatial direction
for the fermion fields,
and the anti-periodic boundary condition is imposed
on the temporal boundary.
We compute lowest 50 eigenpairs at each ``temperature'',
with 48 gauge configurations.
All the eigenvalues $\lambda_{\rm lat}$ of $D$,
which lie on a circle with the radius of $\rho$ in a complex plain,
are stereographically projected onto the imaginary axis
via M{\"o}bius transformation~\cite{Farchioni:1999se},
\begin{equation}
\lambda = \frac{\lambda_{\rm lat}}{1-\lambda_{\rm lat}/2\rho}.
\end{equation}
The spatial volumes are all fixed to $12^3$
and we adopt 5 different temporal lengths,
$N_t=1/T$ = 4, 6, 8, 10, and 12.
The ``finite temperature'' phase transition in compact QED was
extensively investigated~\cite{Vettorazzo:2004cr}
and the ``transition temperature''
at $\beta =1.01$ was found to lie around $N_t\sim 6$.
(See also Ref.~\cite{Panero:2005iu}.)
We expect that our setup can cover the systems 
below and above the transition temperature,
though finite volume effects might not be negligible
and the phase transition temperature would be slightly modified.

\section{Low-lying Dirac eigenvalues and corresponding modes}
\label{diracmodes}

\begin{figure}[h]
\begin{center}
\includegraphics[scale=0.27]{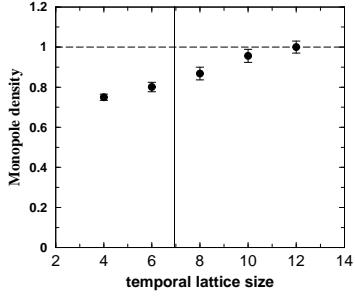}
\end{center}
\caption{\label{monopoledensity}
The normalized monopole densities 
(total monopole lengths divided by total volumes)
are plotted.
The solid line is drawn at $N_t\sim 7$ for reference.
}
\end{figure}

In Ref.~\cite{Takahashi:2007dv}, we found anti-correlations 
between (near-zero) Dirac modes and monopoles,
which indicates that near-zero modes are ``scattered'' by monopoles,
and it was conjectured that monopoles as impurities for near-zero modes
are responsible for the ``complexity'' of a vacuum
and control the level dynamics of low-lying
Dirac eigenvalues~\cite{Takahashi:2007dv}.
Abundance of monopoles implies a complex vacuum
and the level spacing distributions of Dirac eigenvalues
obey the Wigner distribution.
On the other hand, absence of monopoles makes a vacuum simple
and the Poisson-like distribution appears.
The Wigner (Poisson-like) distribution
is the consequence of strong (weak) repulsive forces among
Dirac eigenvalues, and strong (weak) repulsive forces
are responsible for a non-vanishing (vanishing)
spectral density at the spectral origin.
The spectral density at the origin is finally related to
chiral condensate $\langle \bar \psi \psi \rangle$
via the Banks-Casher relation~\cite{Banks:1979yr}.
Chiral condensate $\langle \bar \psi \psi \rangle$
simply reflects the complexity of a vacuum,
which is brought about by monopoles' degrees of freedom.
These results in Ref.~\cite{Takahashi:2007dv} 
were all obtained in isotropic systems
where the temporal length $N_t$ is equal to the spatial extent $N_s$
($N_t = N_s \gg 1/T_c$).
Then, is this scenario the case
also at ``finite temperature'' ($N_t<N_s$)?
The answer would be no.
Though the chirally symmetric vacuum in the weak coupling regime
is actually simple~\cite{Takahashi:2007dv},
we expect that the symmetric vacuum realized
at high temperature is still complex.
In fact, monopole density is not much reduced
even at high temperature (small $N_t$) region,
which can be found in Fig.~\ref{monopoledensity}.

We go further with the clarification
of the chiral phase transition in compact QED.
The keywords are
{\it (1) spectral densities at the spectral origin},
{\it (2) level spacing distributions of Dirac eigenvalues},
{\it (3) correlations between low-lying Dirac modes and monopoles}, and
{\it (4) spatial distributions of low-lying Dirac modes}.

\subsection{Spectral densities and level spacing distributions}

\begin{figure}[h]
\begin{center}
\includegraphics[scale=0.27]{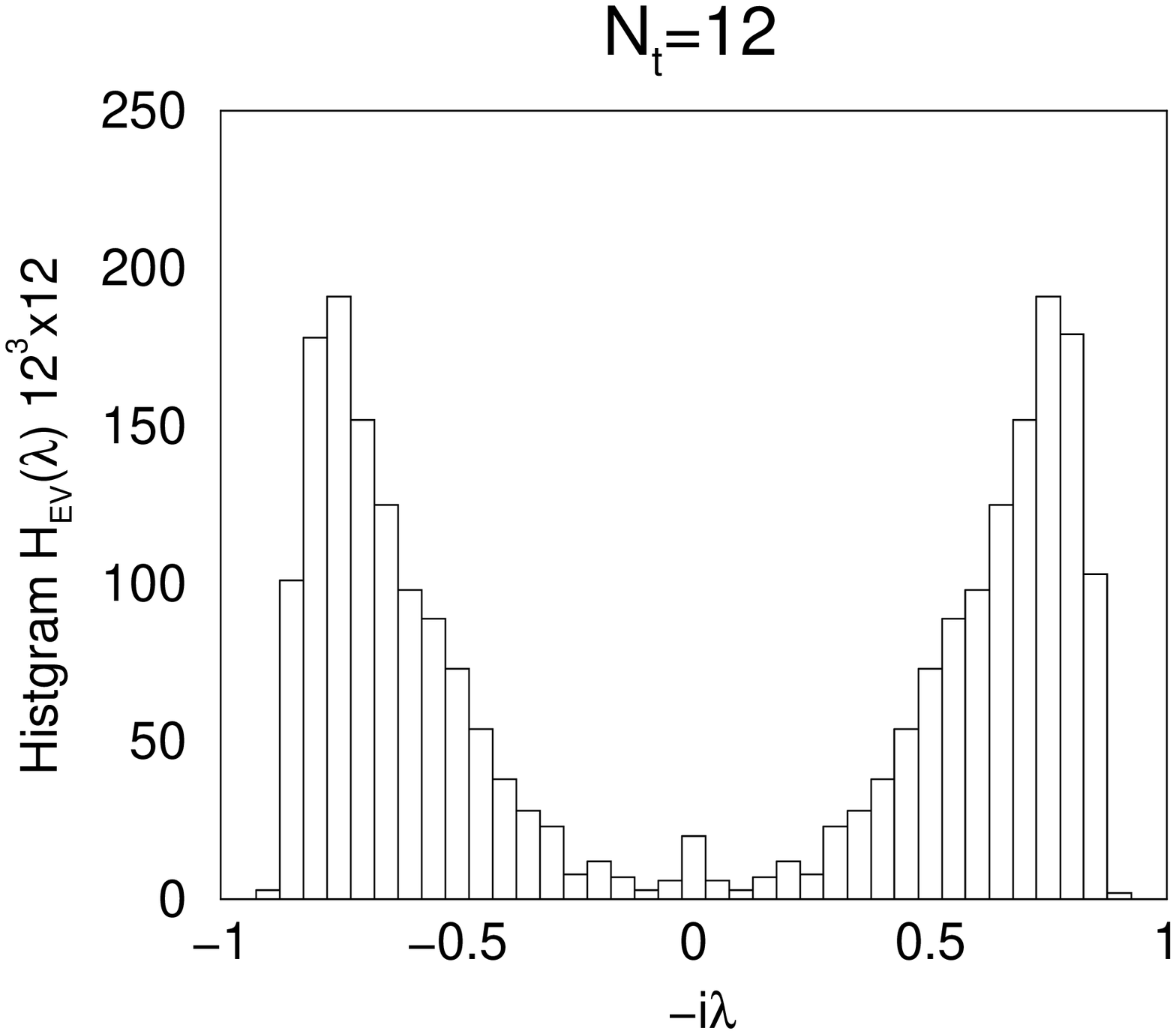}
\includegraphics[scale=0.27]{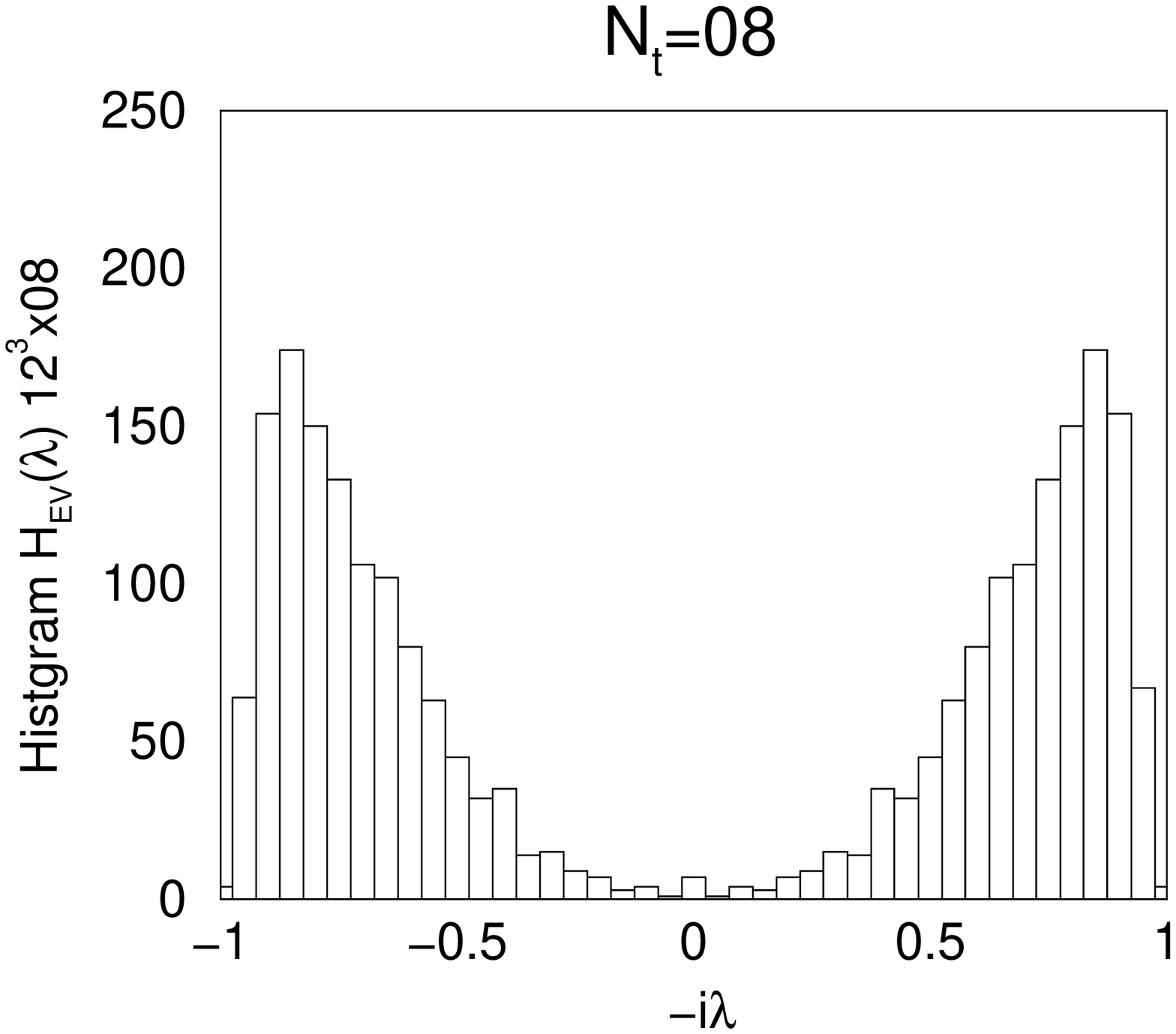}
\includegraphics[scale=0.27]{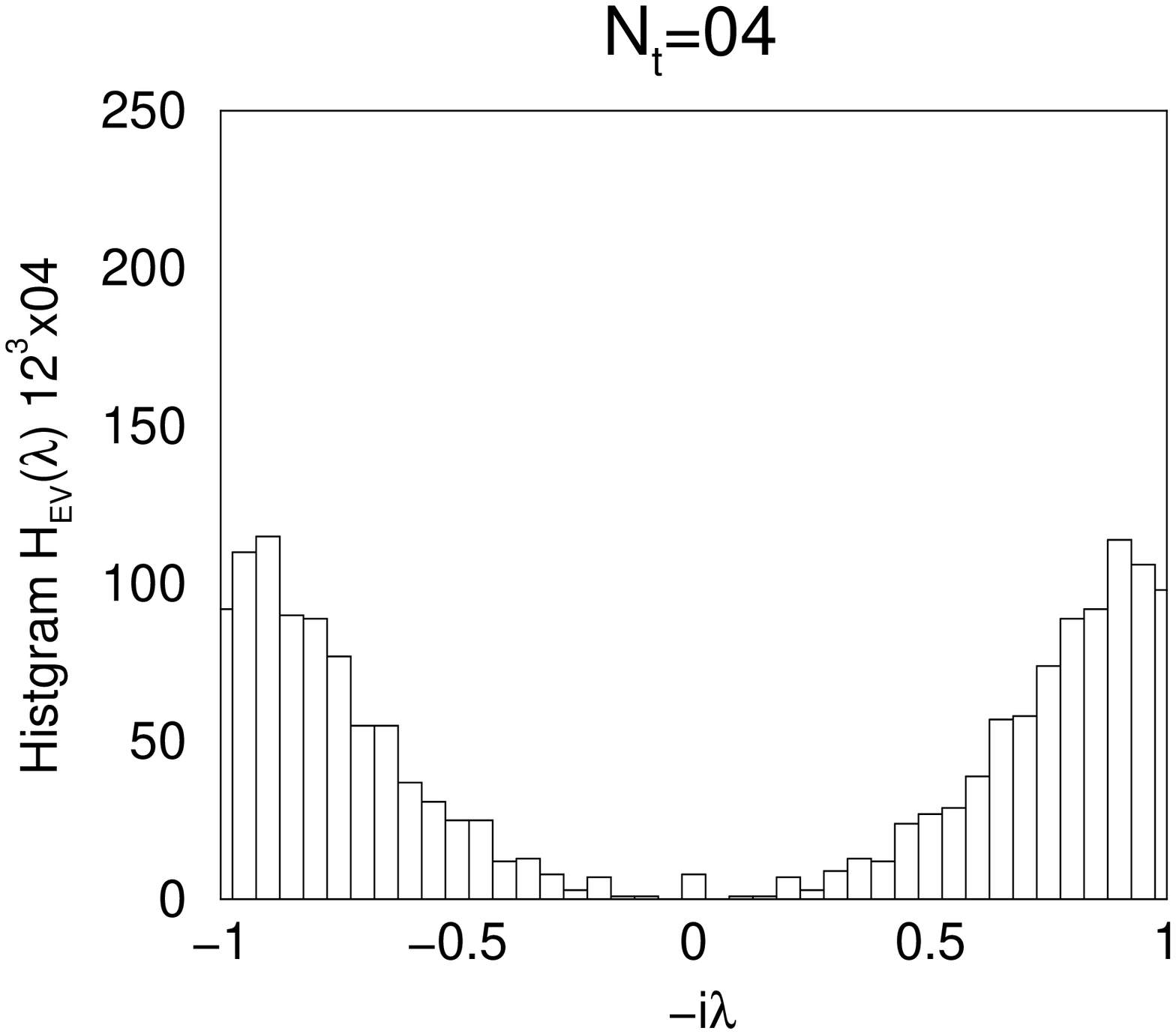}
\end{center}
\caption{\label{histogram}
The histograms $H_{\rm ev}(\lambda)$ of the eigenvalues $\lambda$
of the overlap-Dirac operator $D$ are plotted for each $N_t$(12,8,4).
The horizontal axis denotes $-i\lambda$.
All the eigenvalues $\lambda_{\rm lat}$ lying 
on a circle in a complex plain
are stereographically projected onto the imaginary axis
via M{\"o}bius transformation,
$\lambda = \frac{\lambda_{\rm lat}}{1-\lambda_{\rm lat}/2\rho}$.
}
\end{figure}

We show in Fig.~\ref{histogram} the histograms of Dirac eigenvalues
at $N_t$=12, 8, and 4.
The numbers of zero modes found in 48 gauge configurations at each
inverse temperature $1/T=N_t$ 
(4, 6, 8, 10, 12) are listed in Table~\ref{zeros}.
\begin{table}
\begin{center}
\begin{tabular}[h]{c|ccccc}\hline\hline
$N_t$        & 4  & 6  & 8  & 10  & 12  \\ \hline
$\nu=0$      & 41 & 38 & 41 & 28  & 32  \\
$\nu=1$      & 6  & 10 & 7  & 17  & 12  \\
$\nu=2$      & 1  & 0  & 0  & 3   & 4   \\ \hline\hline
\end{tabular}
\caption{\label{zeros}
The numbers of exact zero-modes found in 48 gauge configurations at each 
inverse temperature $N_t$ are listed. The $i$-th row
gives the number of configurations with 0, 1, 2 zero-mode(s), respectively.
}
\end{center}
\end{table}
The histograms around the spectral origin
at $N_t$=8, 12 are found to be rather flattened and have non-zero heights,
while that at $N_t$=4 exhibits the vanishing density.
These densities are related to
the non-vanishing (vanishing) chiral condensate
at $N_t > 6 \sim 1/T_c$ ($N_t < 6 \sim 1/T_c$) via the Banks-Casher relation.
It is remarkable that the chiral phase transition
is almost accompanied by the deconfinement transition,
which would occur around $N_t \sim 6$.
\begin{figure}[h]
\begin{center}
\includegraphics[scale=0.27]{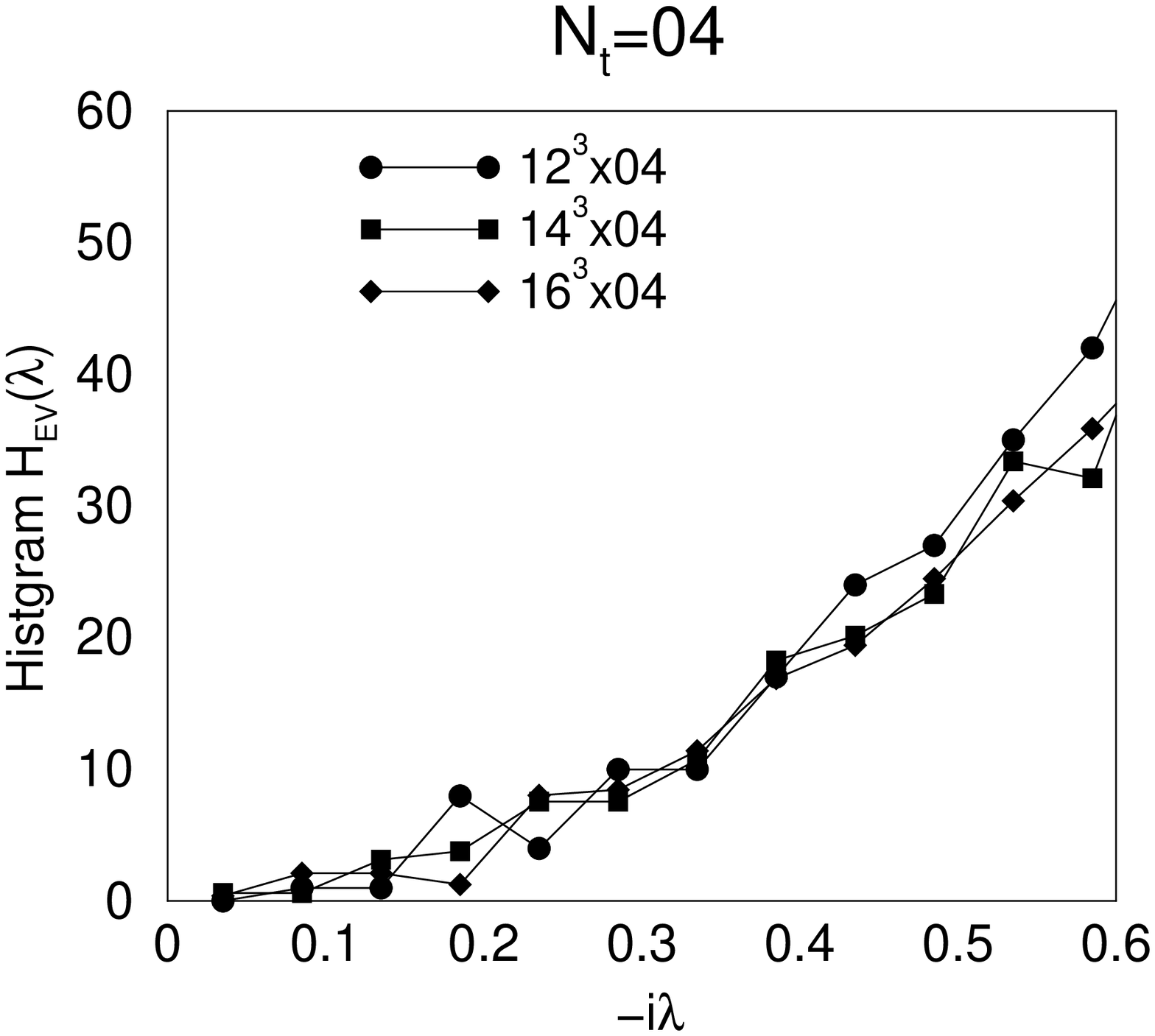}
\includegraphics[scale=0.27]{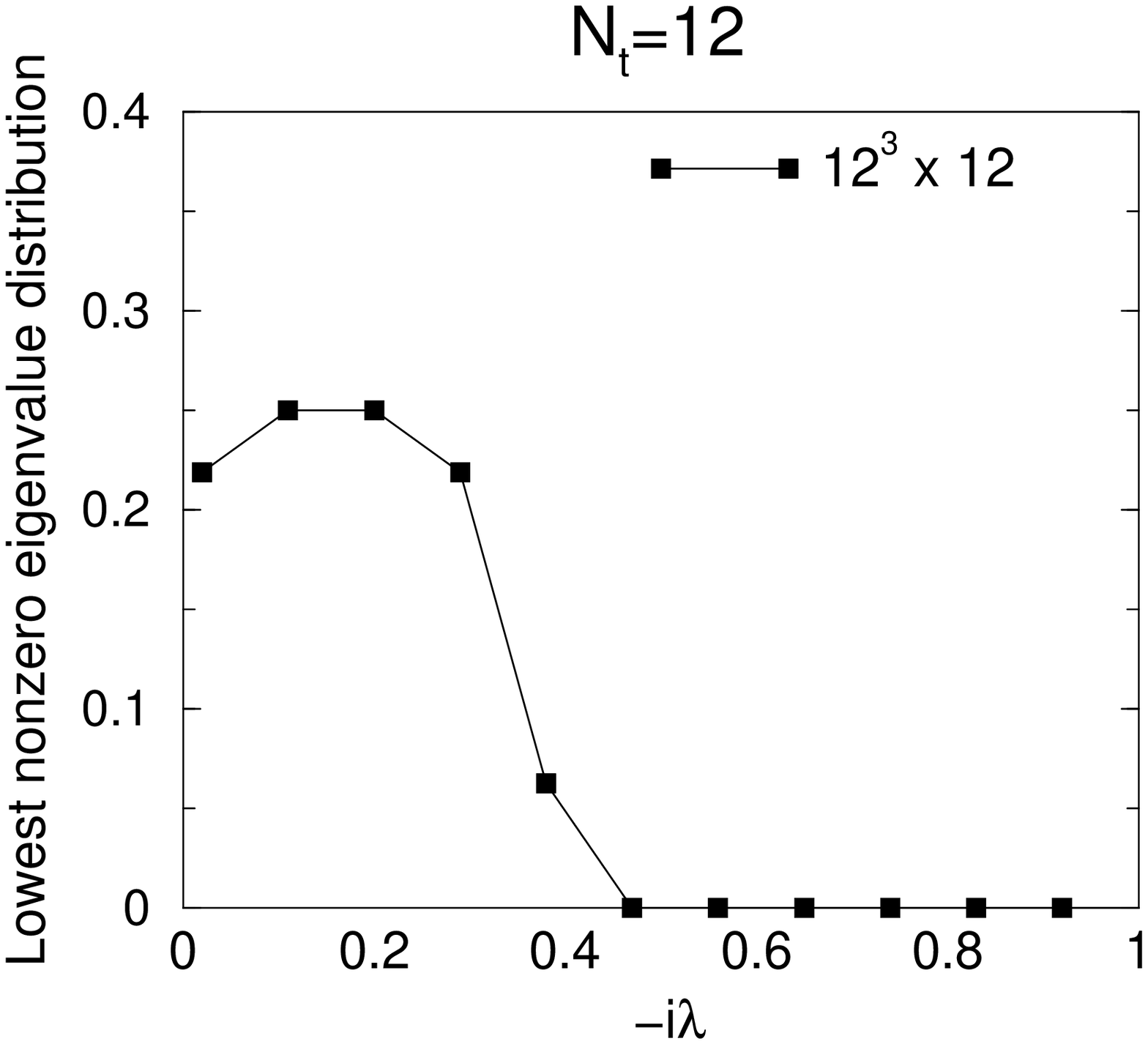}
\includegraphics[scale=0.27]{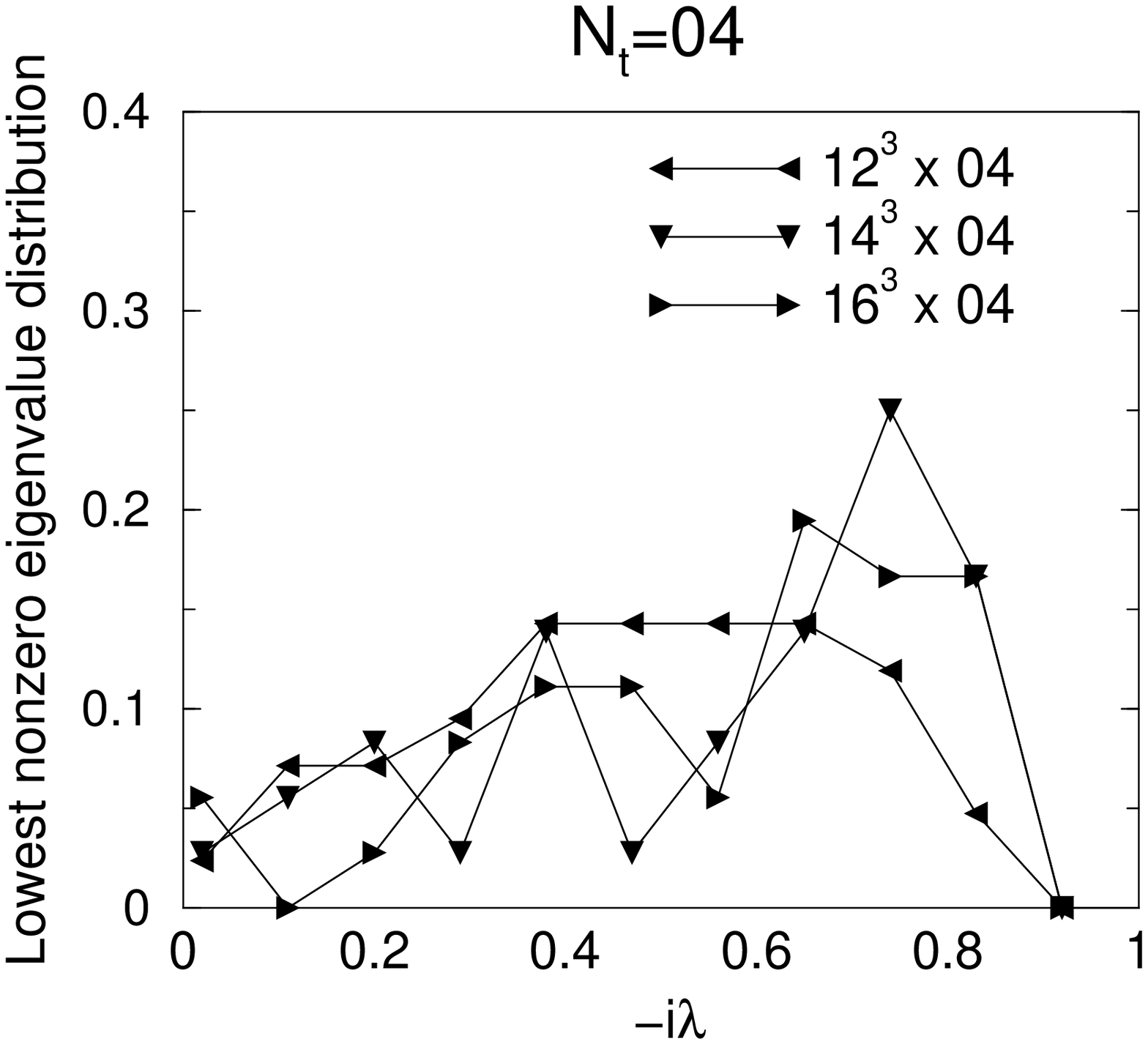}
\end{center}
\caption{\label{histogram2}
{\bf Left:} The histograms $H_{\rm ev}(\lambda)$ of the eigenvalues $\lambda$
of the overlap-Dirac operator $D$, which are
obtained with $12^3\times 4$, $14^3\times 4$, and $16^3\times 4$ lattices.
{\bf Middle and Right:} The distributions of lowest nonzero eigenvalues
at $N_t = 12$ and $4$ obtained with $\nu=0$ gauge configurations.
}
\end{figure}
The vanishing density seen in the $12^3\times 4$ system
could be the consequence of some finite (spatial) volume effects.
In order to ensure that it is not the case,
we show the normalized histograms $H_{\rm ev}(\lambda)$
obtained with $12^3\times 4$, $14^3\times 4$, and $16^3\times 4$ lattices
in Fig.~\ref{histogram2}[left].
All three histograms exhibit the same behavior,
and we can neglect the finite volume effects in $12^3\times 4$ system.
The lowest-nonzero-eigenvalue distributions at $N_t=12$ and $4$
in $\nu=0$ sector
are respectively plotted in Fig.~\ref{histogram2}[middle]
and Fig.~\ref{histogram2}[right], for the purpose of reference.

\begin{figure}[h]
\begin{center}
\includegraphics[scale=0.27]{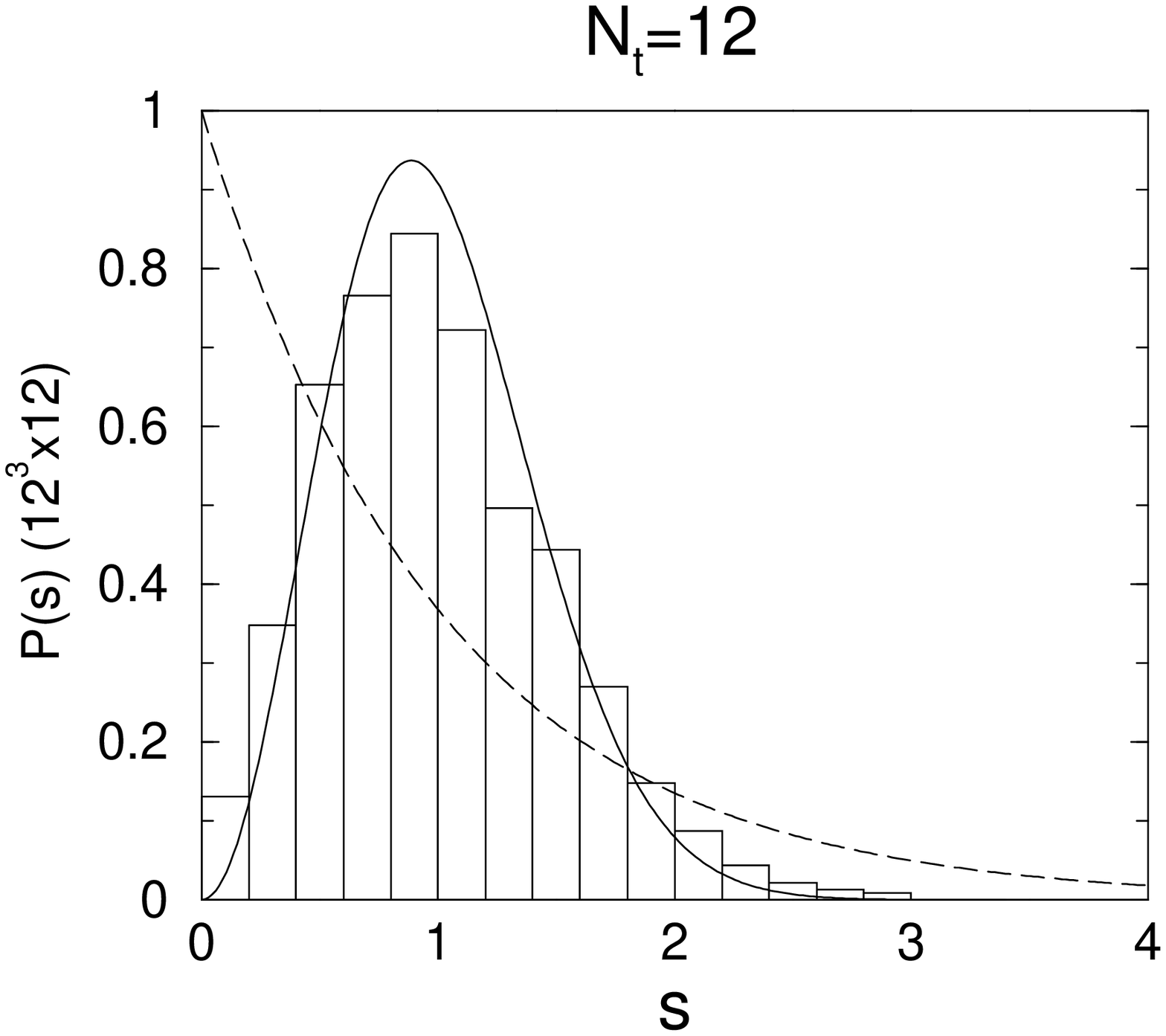}
\includegraphics[scale=0.27]{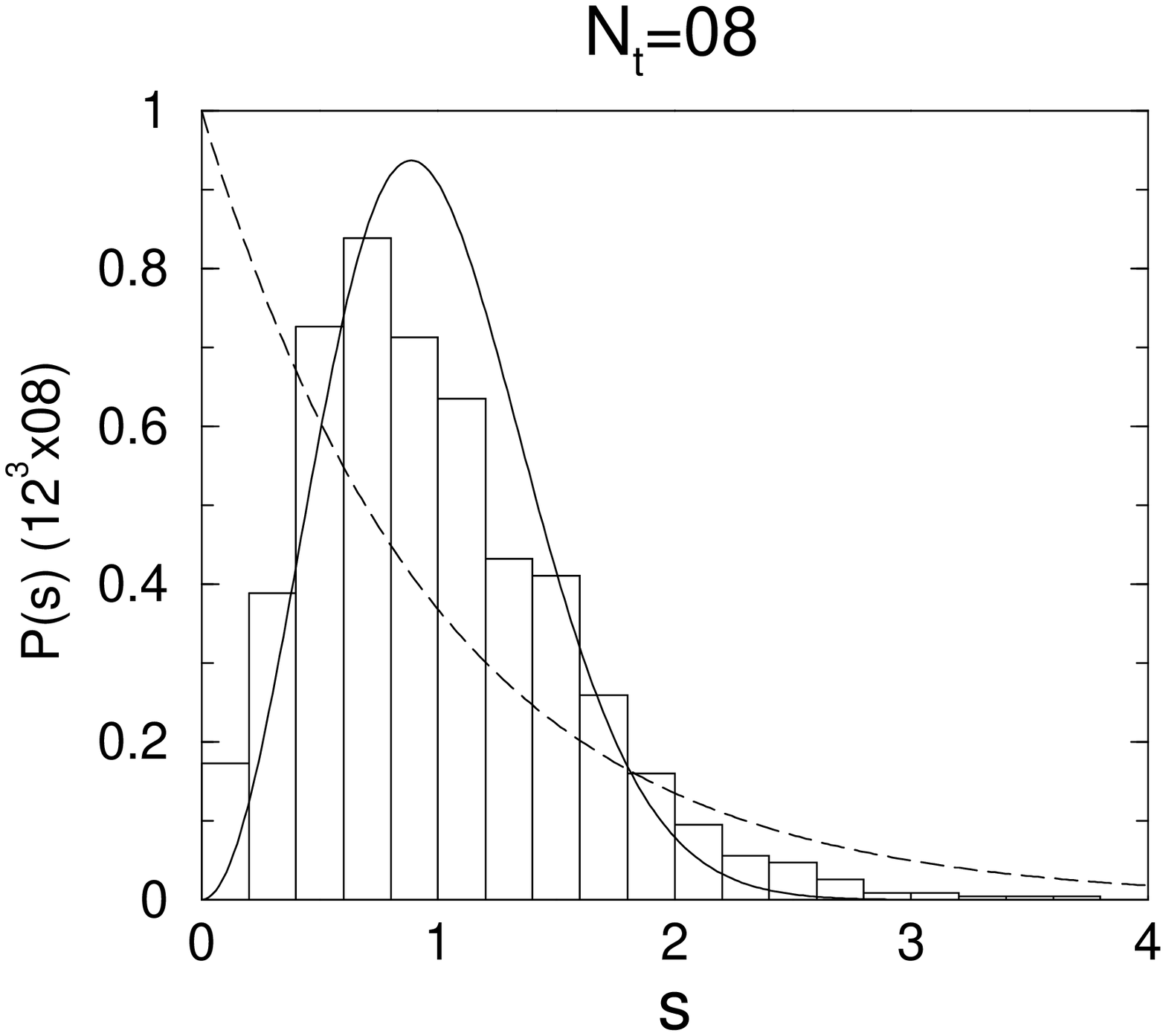}
\includegraphics[scale=0.27]{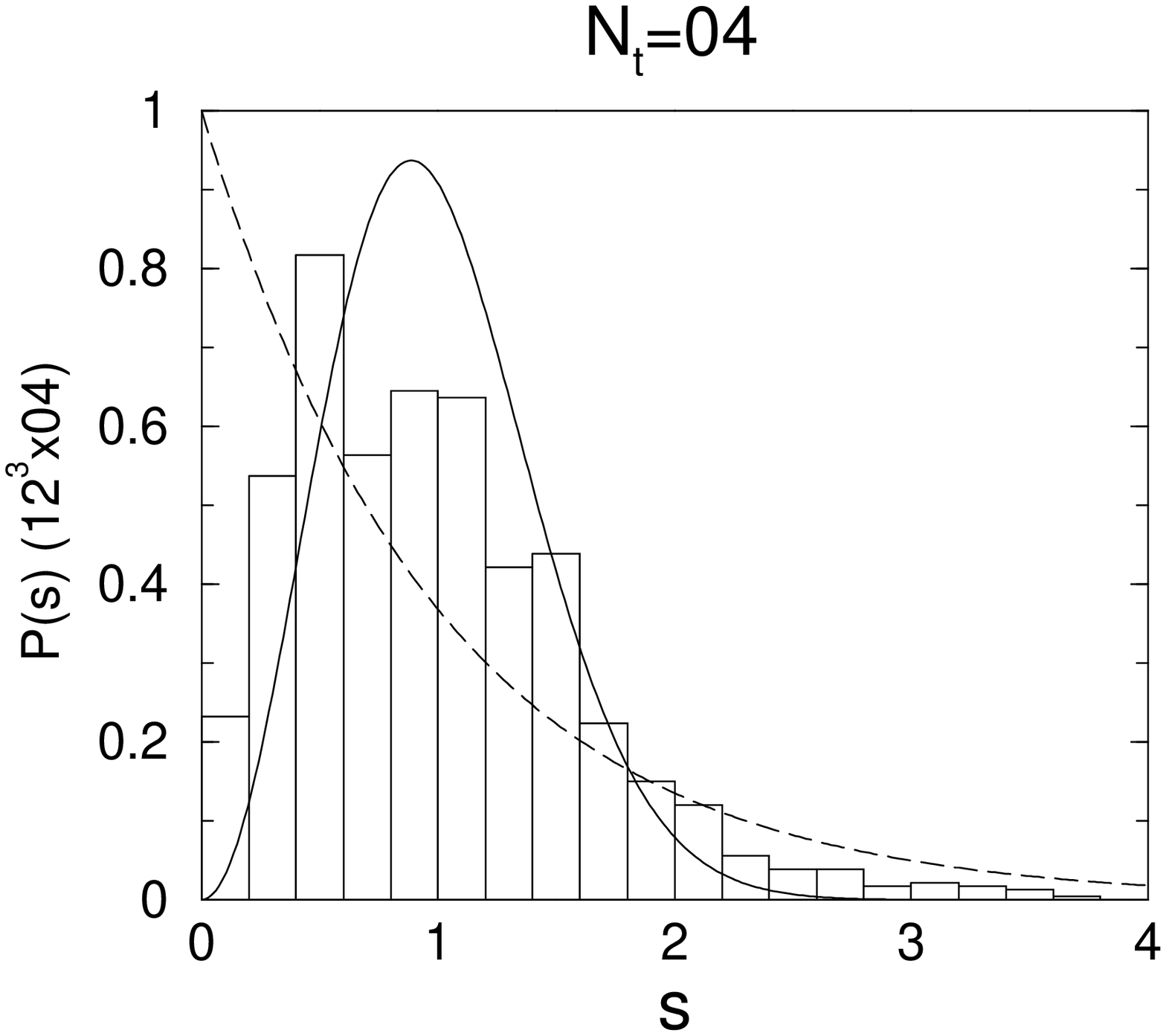}
\end{center}
\caption{\label{levelstat}
The unfolded nearest-neighbor level spacing distributions $P_{\rm lat}(s)$
at $N_t$=12, 8, and 4.
The solid lines denote the Wigner distribution function
$P_{\rm Wig}(s)\equiv\frac{32}{\pi^2}s^2\exp (-\frac{4}{\pi}s^2)$
and the dashed lines the Poisson distribution function
$P_{\rm Poi}(s)=\exp (-s)$.
}
\end{figure}

As actually demonstrated in Refs.~\cite{Takahashi:2007dv,Garcia-Garcia:2006gr},
the spectral density at the spectral origin is enhanced by
the ``repulsive force'' among eigenvalues,
and this force can be clarified with
the neighboring level spacing distributions of Dirac eigenvalues.
We show in Fig.~\ref{levelstat} 
the unfolded neighboring level spacing distributions
$P_{\rm lat}(s)$
obtained at $N_t$=12, 8, and 4.
The solid lines denote the Wigner distribution function
$P_{\rm Wig}(s)\equiv\frac{32}{\pi^2}s^2\exp (-\frac{4}{\pi}s^2)$,
which is a good approximation of the original distribution
obtained by the random matrix theory
with chiral unitary ensemble,
and the dashed lines the Poisson distribution function
$P_{\rm Poi}(s)=\exp (-s)$.
We again find the Wigner-Poisson transition,
which was also found in Ref.~\cite{Takahashi:2007dv}
and is consistent with the spectral densities at the origin.

\subsection{low-lying Dirac modes and monopoles}

\begin{figure}[h]
\begin{center}
\includegraphics[scale=0.27]{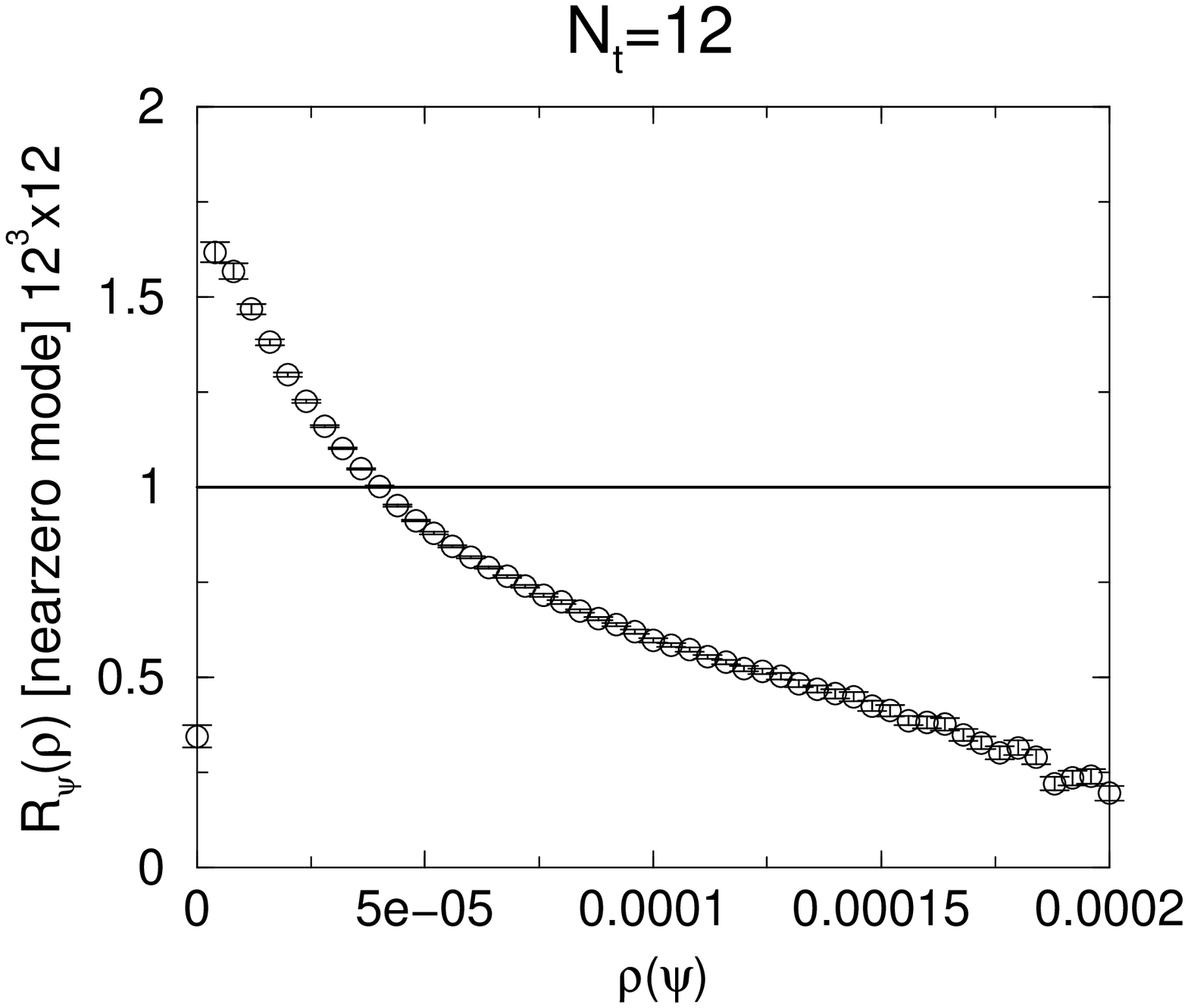}
\includegraphics[scale=0.27]{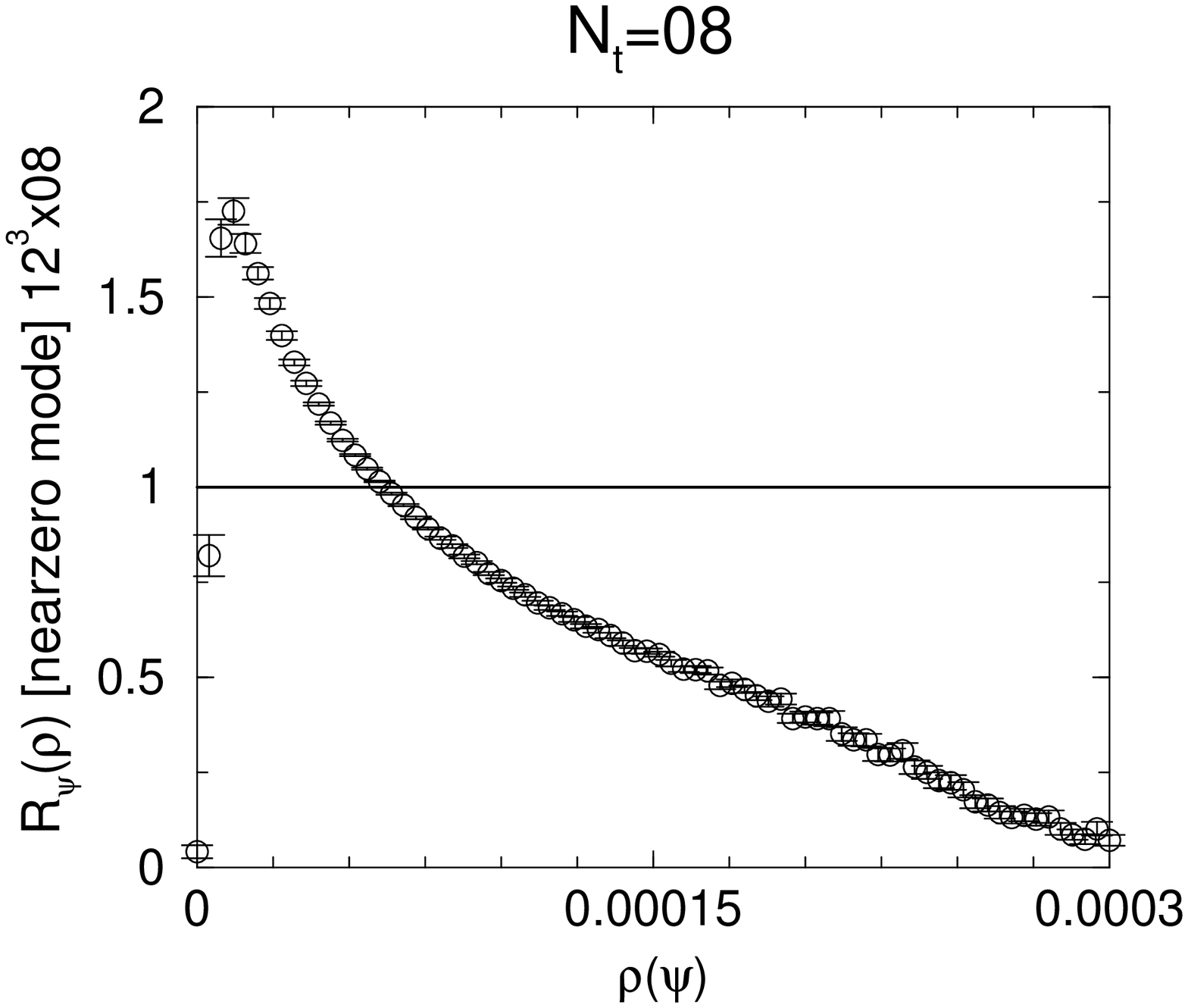}
\includegraphics[scale=0.27]{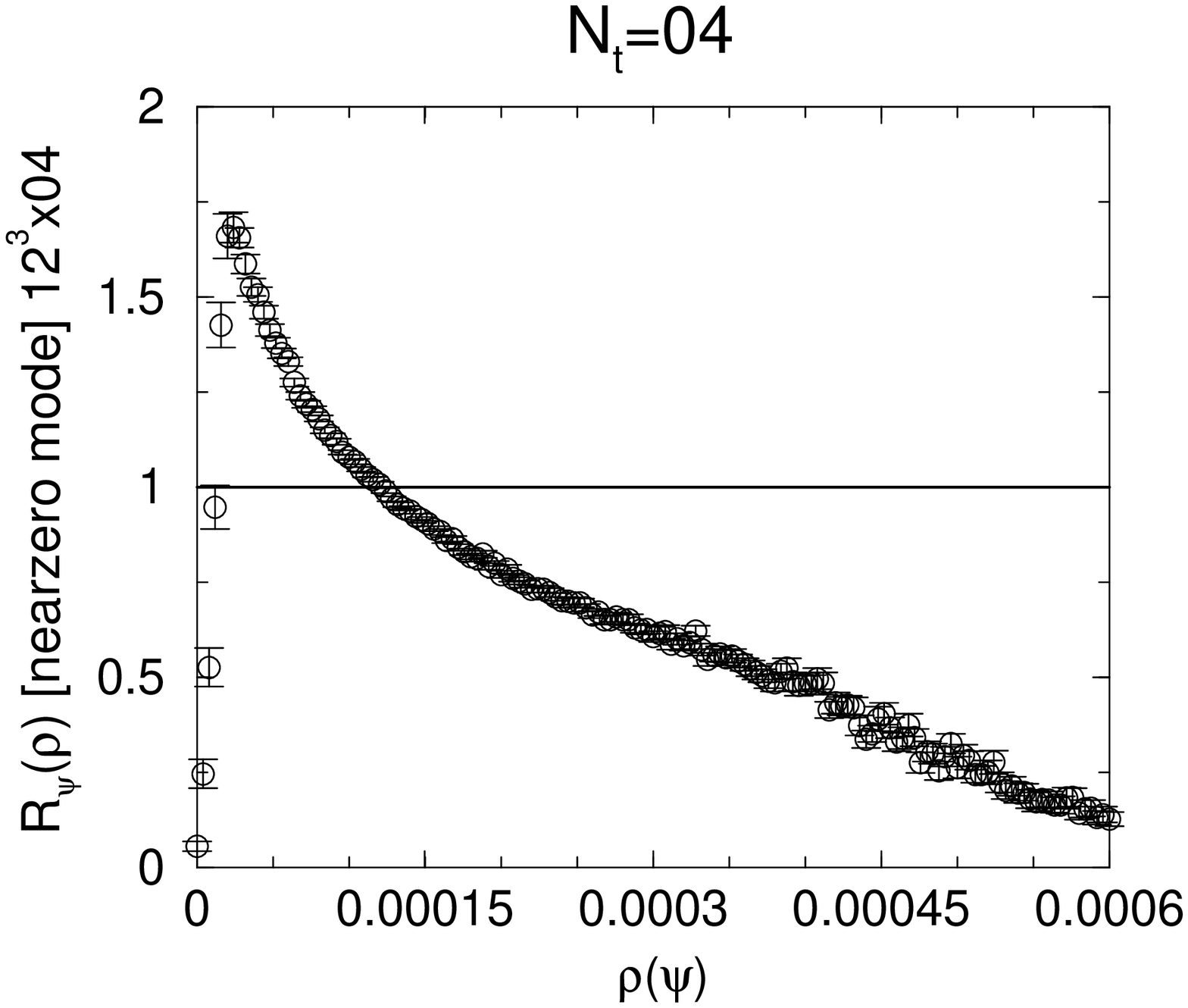}
\end{center}
\caption{\label{ratios2}
The histogram ratios $R_\psi(\rho_\psi)$
for near-zero modes
at $N_t$=12, 8, and 4
are plotted in the range of $0 \leq \rho_\psi \leq 0.0002\times \frac{12}{N_t}$.
We draw a line at $R_\psi(\rho_\psi)=1$ for reference.
}
\end{figure}

We next investigate the correlations between low-lying modes and monopoles
following Ref.~\cite{Takahashi:2007dv}.
For this aim, we define and investigate the histogram ratios
$R_\psi(\rho_\psi)$~\cite{Takahashi:2007dv} defined as
\begin{eqnarray}
R_\psi(\rho_\psi)\equiv
\frac
{H^{\rm mon}_{\rm \psi}(\rho_\psi)}
{H^{\rm all}_{\rm \psi}(\rho_\psi)}.
\end{eqnarray}
The eigenmode density $\rho_{\psi}\equiv \sum_{\alpha}|\psi_\lambda(x)|^2$ 
here is the absolute squares of an eigenfunction $\psi_\lambda(x)$
locally summed up over the spinor index.
$H^{\rm all}_{\rm \psi}(\rho_\psi)$ ($H^{\rm mon}_{\rm \psi}(\rho_\psi)$)
denotes the histogram of $\rho_\psi$
evaluated at all the sites (only on monopoles).
This quantity $R_\psi(\rho_\psi)$ equals to 1,
{\it if there is no correlation between the spatial fluctuations of Dirac
modes and monopoles}.
In the case when a positive (negative) correlation 
exists between the spatial fluctuations of Dirac modes and monopoles,
$R_\psi(\rho_\psi) > 1$ at smaller (larger) $\rho_\psi$
and
$R_\psi(\rho_\psi) < 1$ at large (smaller) $\rho_\psi$ hold.
We show in Fig.~\ref{ratios2} the histogram ratios
for near-zero modes obtained at $N_t$=12, 8, and 4.
Display ranges are chosen
as $0 \leq \rho_\psi \leq 0.0002\times \frac{12}{N_t}$
with intrinsic densities in mind.
The universal anti-correlations are again observed.
(We here omit the results on exact-zero 
modes~\cite{Berg:2001nn,Drescher:2004st},
since they show no apparent correlation with monopoles.
This tendency was also found in our previous paper~\cite{Takahashi:2007dv}.
The present strategy is rather simple
and we would need more sophisticated analyses
to cast light on the properties of exact-zero modes.
In the following sections,
attentions will be paid mainly to near-zero Dirac modes.)


\subsection{Spatial distributions of low-lying Dirac modes}

To clarify the spatial distributions of Dirac eigenmodes,
it is convenient to extract the inverse participation ratio (IPR) 
of each eigenmode.
The eigenmode's distribution is partially encoded in the corresponding IPR:
The IPR is unity when $\psi_\lambda(x)$ maximally spreads over the system,
and equals to $V$ in the case when $\psi_\lambda(x)$ 
lives only on a single site,
reflecting the spatial distribution of the eigenfunction $\psi_\lambda(x)$.

The IPR $I(\lambda)$ of $\psi_\lambda(x)$ is defined as 
\begin{equation}
I(\lambda) = V\sum_x \rho_{\rm IPA}(x)^2,\quad
\rho_{\rm IPR}(x)\equiv \sum_{\alpha}|\psi_\lambda(x)|^2.
\end{equation}
Here, $V$ denotes the system volume and
$\psi_\lambda(x)$ is the eigenfunction 
associated with an eigenvalue $\lambda$
normalized as $\sum_x|\psi_\lambda(x)|^2=1$.
The Greek alphabet $\alpha$ is the index for a spinor.
The density $\rho_{\rm IPR}(x)$
is obtained by locally summing up the absolute square of each component 
of an eigenfunction $\psi_\lambda(x)$ only over its spinor index.

\begin{figure}[h]
\begin{center}
\includegraphics[scale=0.27]{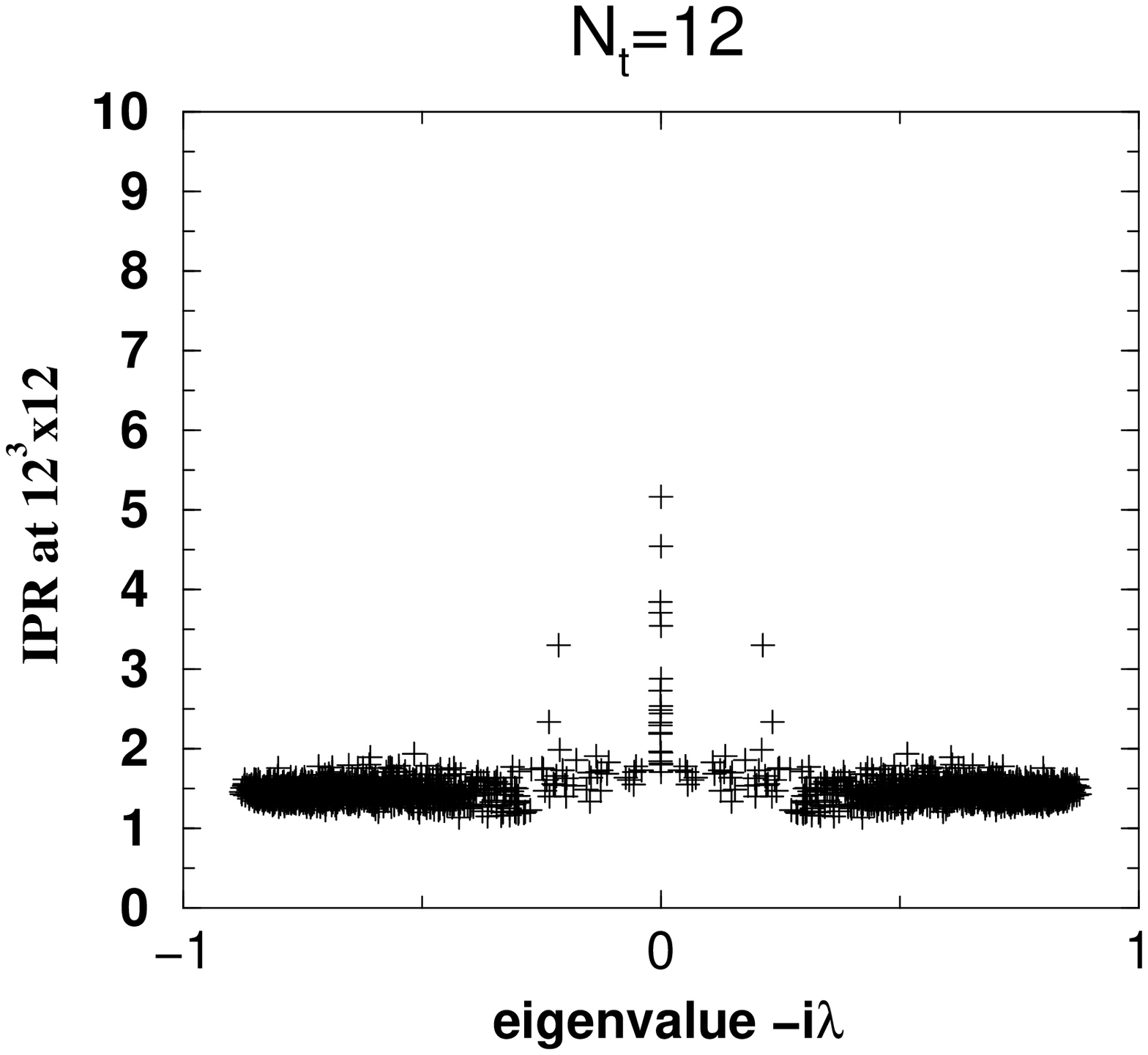}
\includegraphics[scale=0.27]{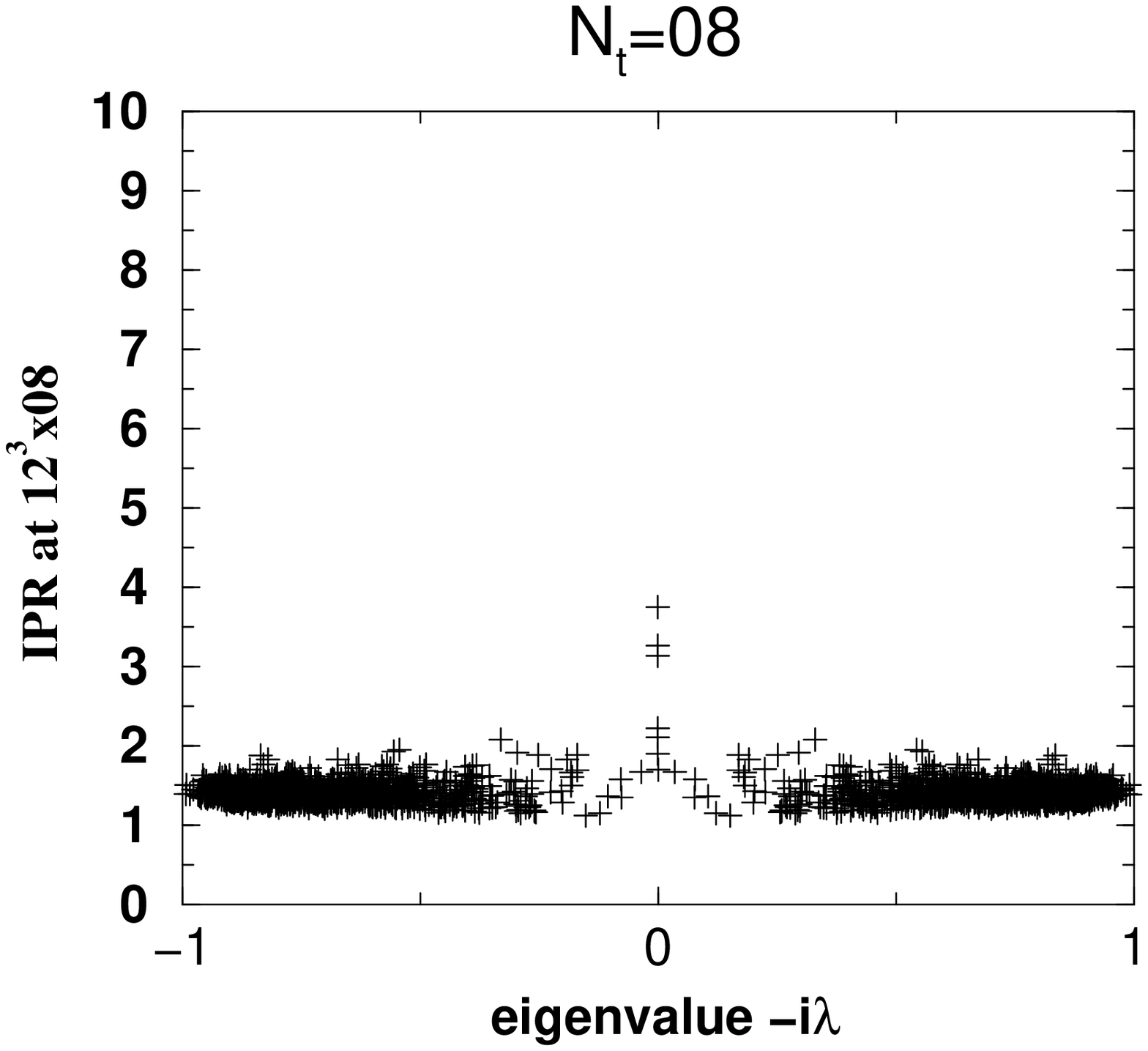}
\includegraphics[scale=0.27]{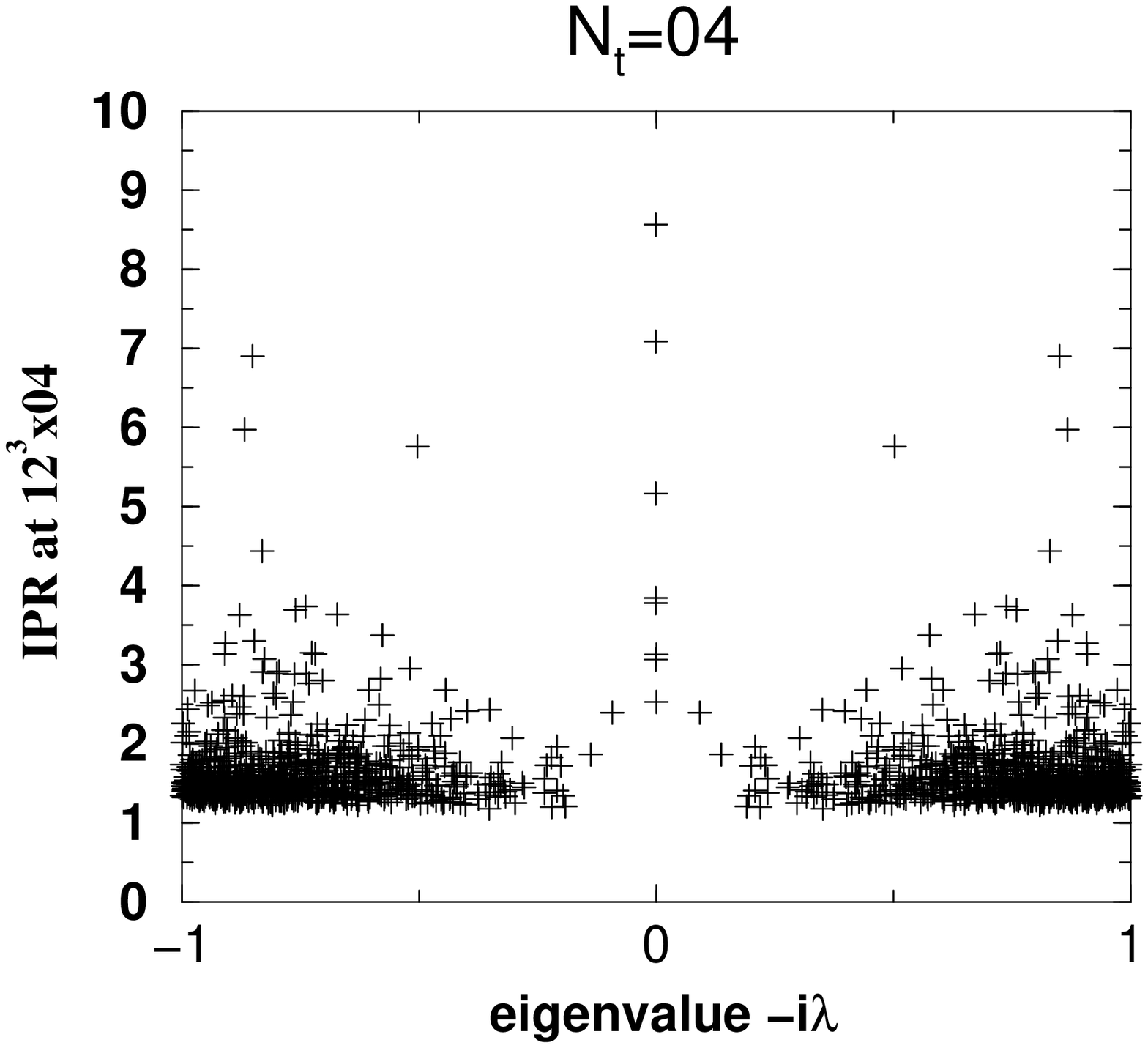}
\end{center}
\caption{\label{ipr}
The scatter plots of 
inverse participation ratios of low-lying Dirac modes
at $N_t$=12, 8, and 4.
The horizontal axis denotes $-i\lambda$,
the associated eigenvalue to each Dirac mode.
}
\end{figure}

We show the scatter plots of the IPRs of low-lying modes
in Fig.~\ref{ipr}.
The IPRs of near-zero modes approximately range from 1 to 2
at $N_t$=8 and 12.
On the other hand, at $N_t$=4,
many of the IPRs of near-zero modes take much larger values
indicating the strong localizations
of near-zero modes above $T_c$.
Near-zero modes are localized and the overlaps among them get much smaller,
which would be the source of the Poisson-like distribution
in the neighboring level spacings of Dirac eigenvalues:
Level repulsions would not occur
without overlaps among eigenfunctions.
Weaker repulsive forces among eigenvalues
characterized by such a Poisson-like statistics lead to
the vanishing eigenvalue density at the spectral origin,
and the chiral symmetry is restored.
The chiral phase transition across $T_c$
is now found to be essentially 
different from that across $\beta_c$~\cite{Takahashi:2007dv}.
The Poisson-like statistics above $\beta_c$,
where near-zero Dirac modes are completely delocalized 
and are plane-wave like, is caused by the ``simpleness'' of the vacuum.
That above $T_c$ is however driven by the localization of near-zero modes.
We expect nontrivial vacuum structures above $T_c$.

\begin{figure}[h]
\begin{center}
\includegraphics[scale=0.27]{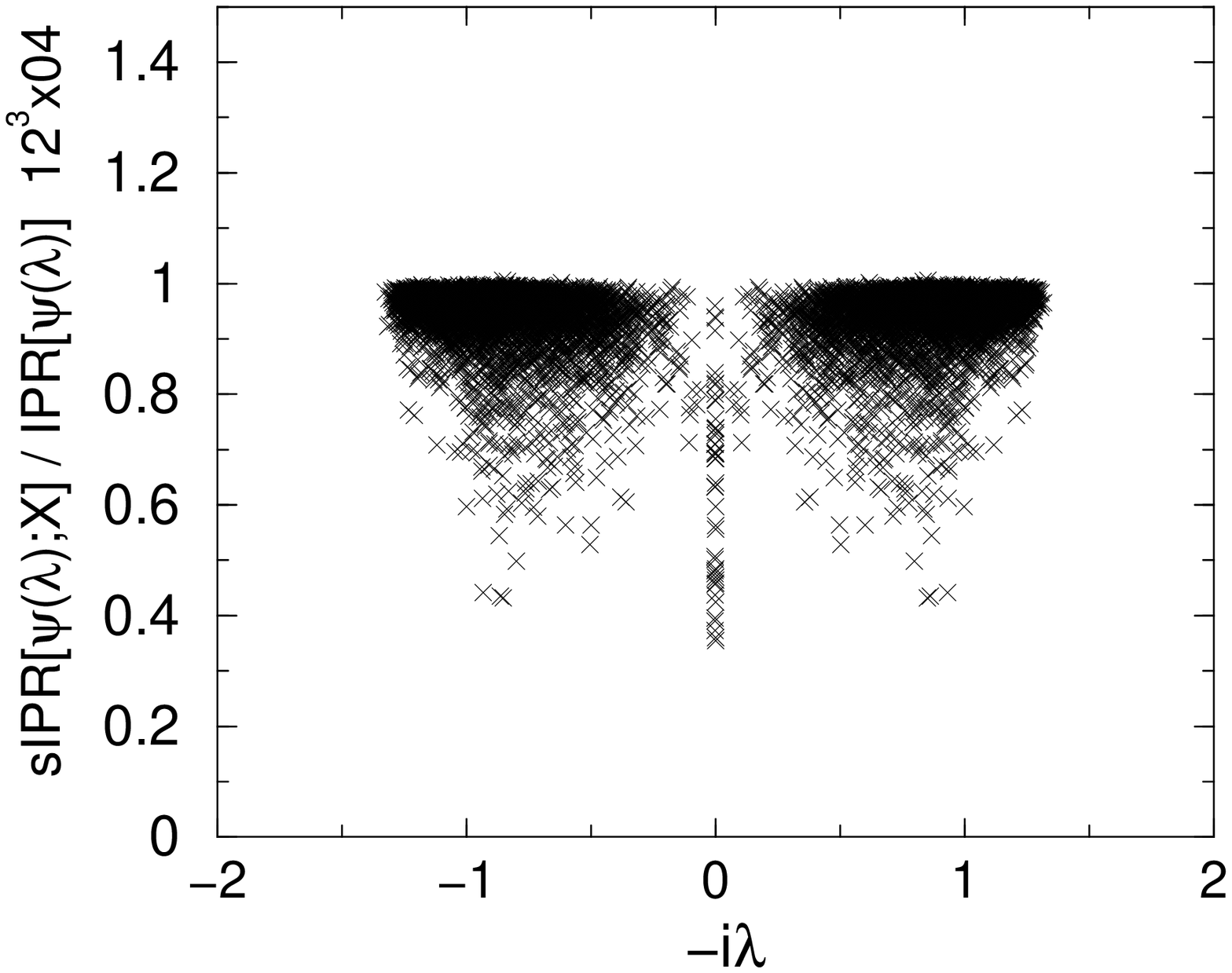}
\includegraphics[scale=0.27]{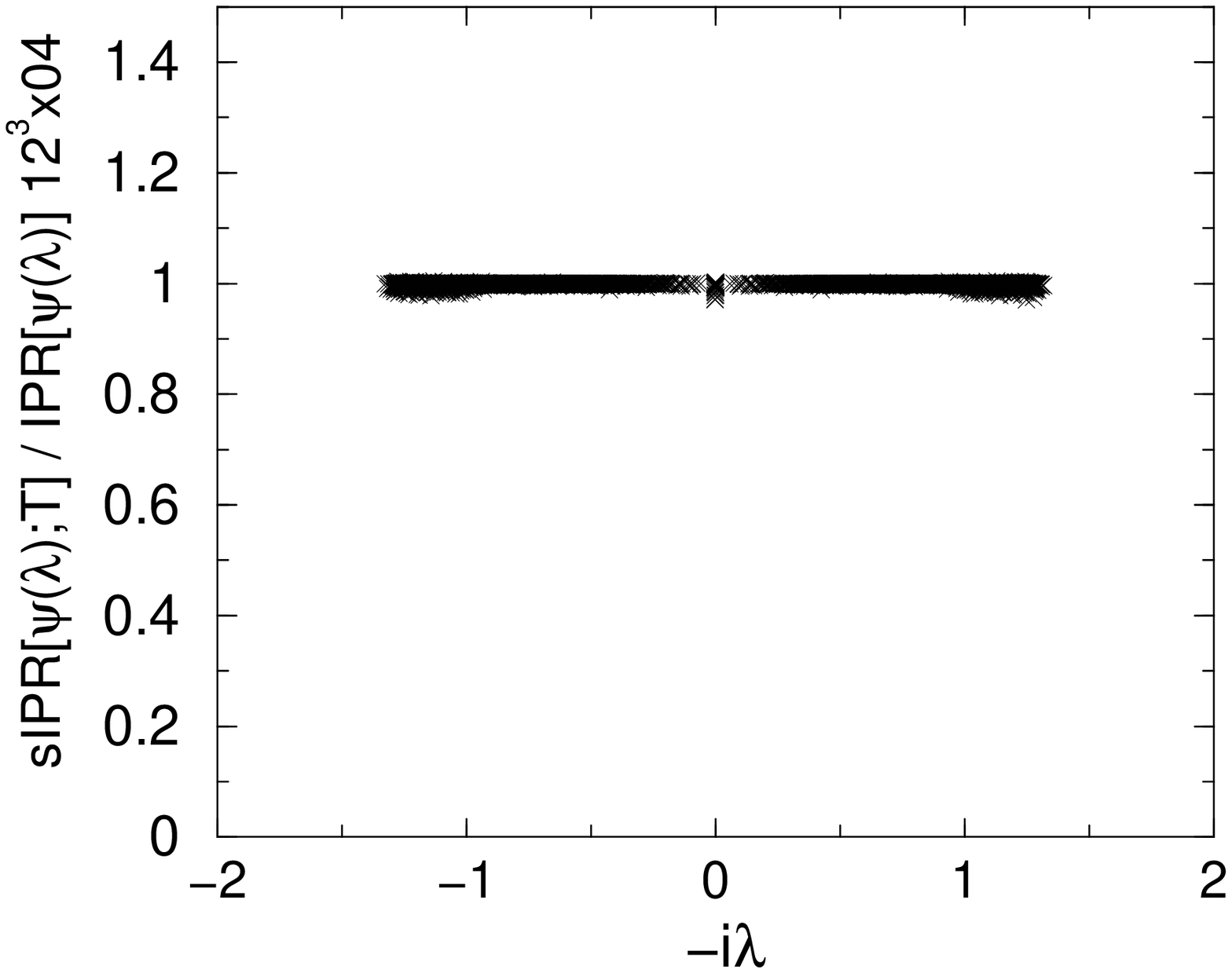}
\end{center}
\caption{\label{iprratio1}
The IPR-ratios, $Is(\lambda ; i)/I(\lambda) (i= 1,2,3)$ {[left]}
and
$Is(\lambda ; 4)/I(\lambda)$ {[right]}
obtained with the $12^3\times 4$ lattice are plotted.
}
\end{figure}

\begin{figure}[h]
\begin{center}
\includegraphics[scale=0.4]{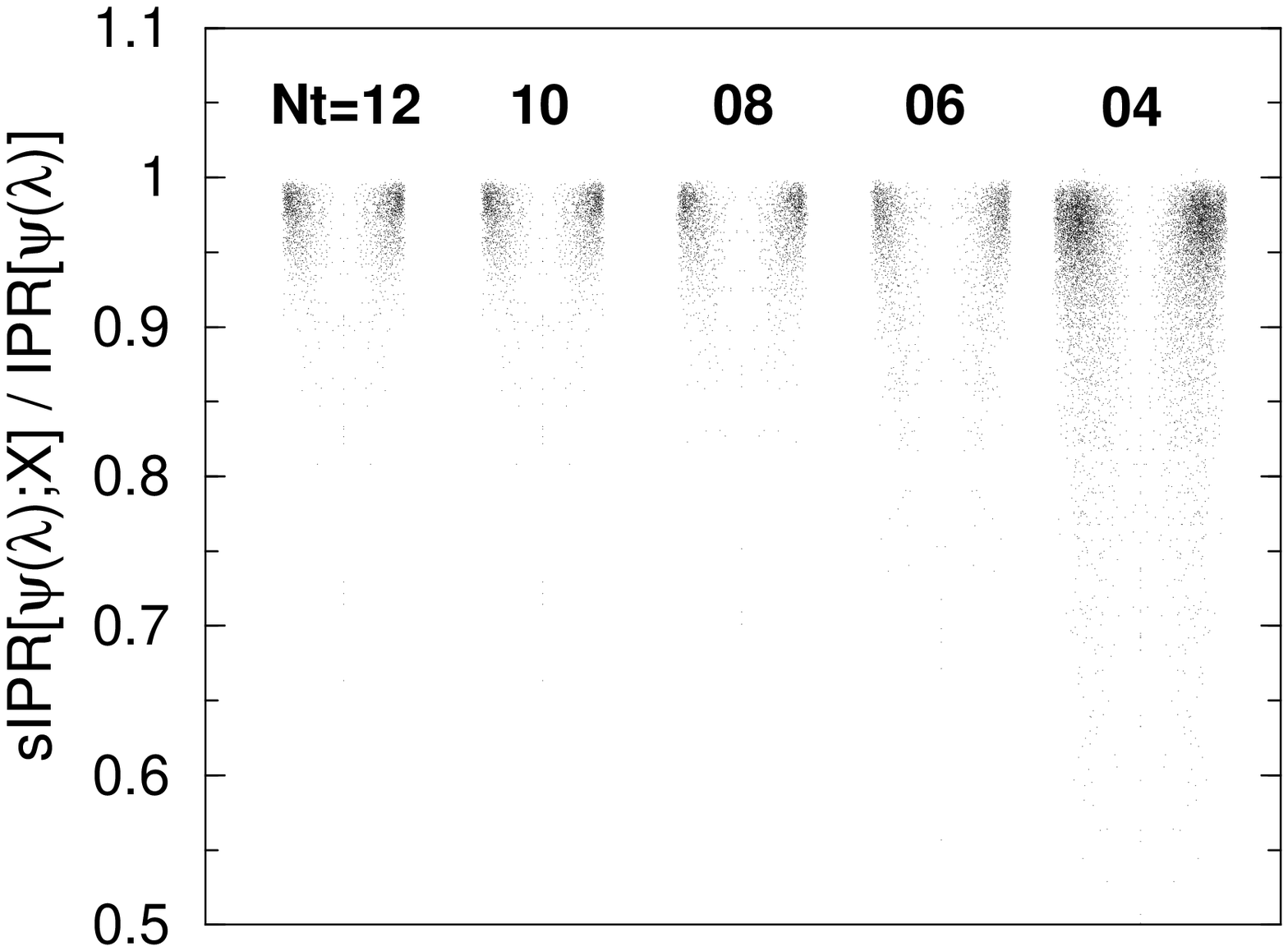}
\includegraphics[scale=0.4]{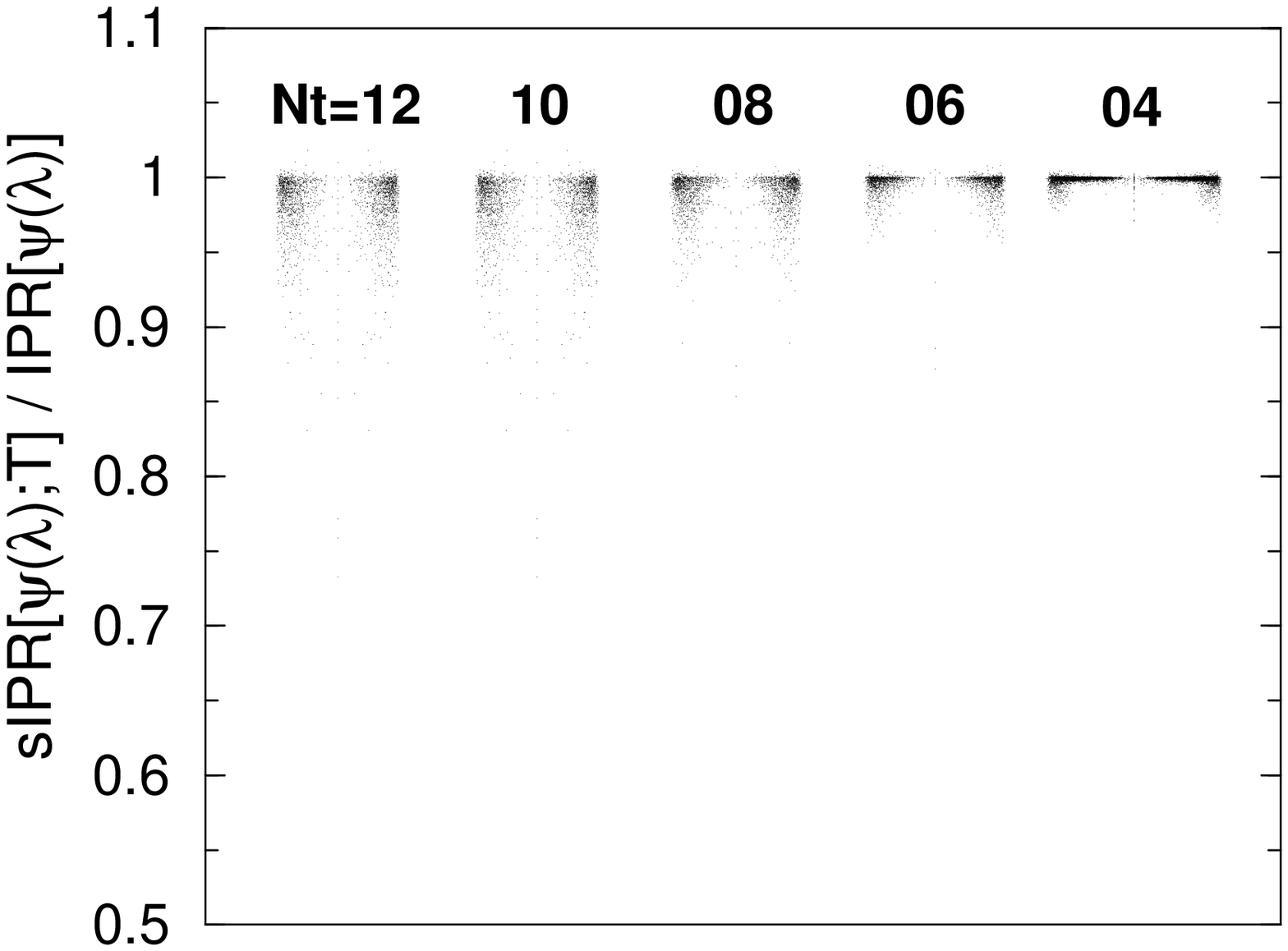}
\end{center}
\caption{\label{iprratio2}
The IPR-ratios, $Is(\lambda ; i)/I(\lambda) (i= 1,2,3)$ {[left]} and
$Is(\lambda ; 4)/I(\lambda)$ {[right]}
obtained with $12^3\times 4\sim 12^3\times 12$ lattice are plotted.
}
\end{figure}

We define another type of IPR $Is(\lambda;\mu)$,
which is nothing but 
the IPR evaluated in a 3-dim subspace at $x_\mu = const.$,
for further clarification of Dirac modes' distributions above $T_c$.
The subspace-IPR $Is(\lambda;\mu)$ is defined as,
\begin{equation}
Is(\lambda;\mu) \equiv
V_s \sum_{x_\mu =const.} \rho_{\rm IPA}'(x;\mu)^2,\quad
\rho_{\rm IPR}'(x;\mu)\equiv 
\sum_{\alpha}|\psi_\lambda(x)|^2
/
\sum_{\alpha,x,x_\mu =const.}|\psi_\lambda(x)|^2.
\end{equation}
Here, $V_s$ denotes the volume of the 3-dim subspace.
We evaluate the ratios,
$Is(\lambda;i)/I(\lambda)(i= 1,2,3)$ 
and
$Is(\lambda;4)/I(\lambda)$.
In case an eigenfunction is simply extended along the temporal direction
and spatially localized,
$Is(\lambda;4)/I(\lambda)$ is just unity
and
$Is(\lambda;i)/I(\lambda)(i= 1,2,3)$ is expected to be less than 1,
because in this case the eigenfunction has no structure
along the time-direction and only the spatial structure
is responsible for the total IPR.
We show in Fig.~\ref{iprratio1}
$Is(\lambda;4)/I(\lambda)$ and
$Is(\lambda;i)/I(\lambda)(i= 1,2,3)$
obtained in $12^3\times 4$ lattice.
One can find that $Is(\lambda;4)/I(\lambda)\simeq 1$
and $Is(\lambda;i)/I(\lambda)(i= 1,2,3) \ll 1$ hold,
which indicates the vanishing temporal structure 
of low-lying Dirac modes at $N_t=4$.
All the ratios,
$Is(\lambda;4)/I(\lambda)$ and
$Is(\lambda;i)/I(\lambda)(i= 1,2,3)$,
evaluated at $4\le N_t\le 12$ are shown in Fig.~\ref{iprratio2}.
The temporal structures of low-lying modes
quickly vanish around $N_t\sim 6\sim 1/Tc$.

\section{Discussions}
\label{discussions}

Now that we have found that, at ``high temperature'',
low-lying (near-zero) Dirac modes lose their temporal structures
and are spatially localized.
We have two questions:
{\it (1)Why do the temporal structures of near-zero modes vanish?}
{\it (2)Why are near-zero modes spatially (and strongly) localized?}
As we have just found or as can be found in Ref.~\cite{Takahashi:2007dv},
near-zero Dirac modes have anti-correlations with monopoles and then
monopoles are surely responsible for the distributions of near-zero modes.
These features may come from the metamorphosis of monopole world lines.

\begin{figure}[h]
\begin{center}
\includegraphics[scale=0.22]{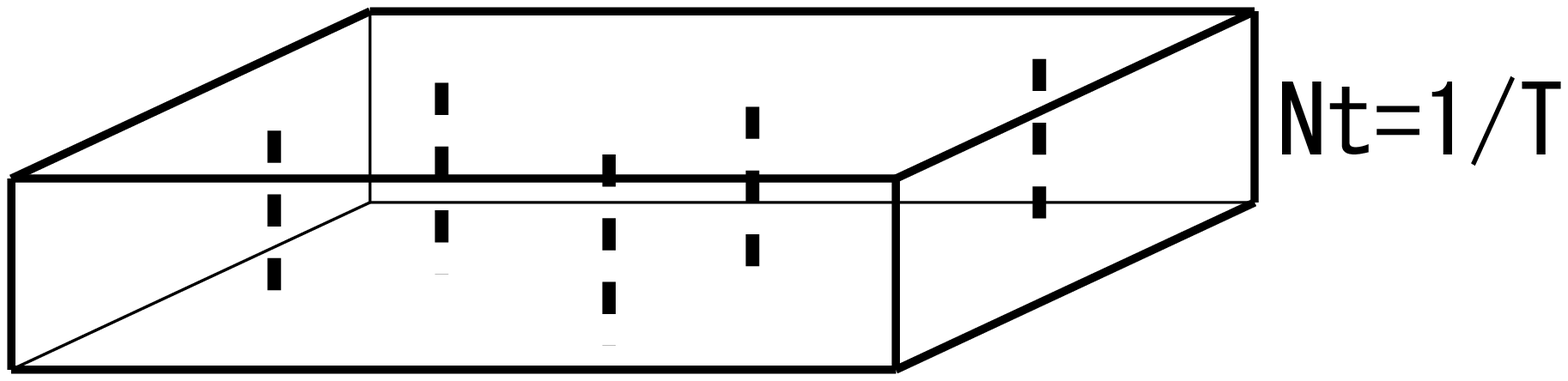}
\hspace{.2cm}
\includegraphics[scale=0.22]{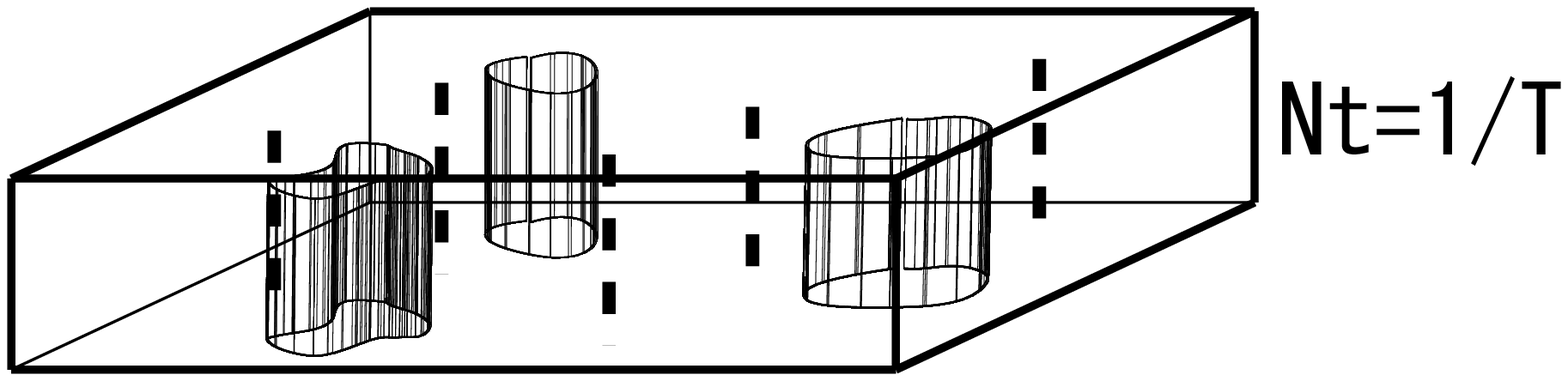}
\hspace{.2cm}
\includegraphics[scale=0.22]{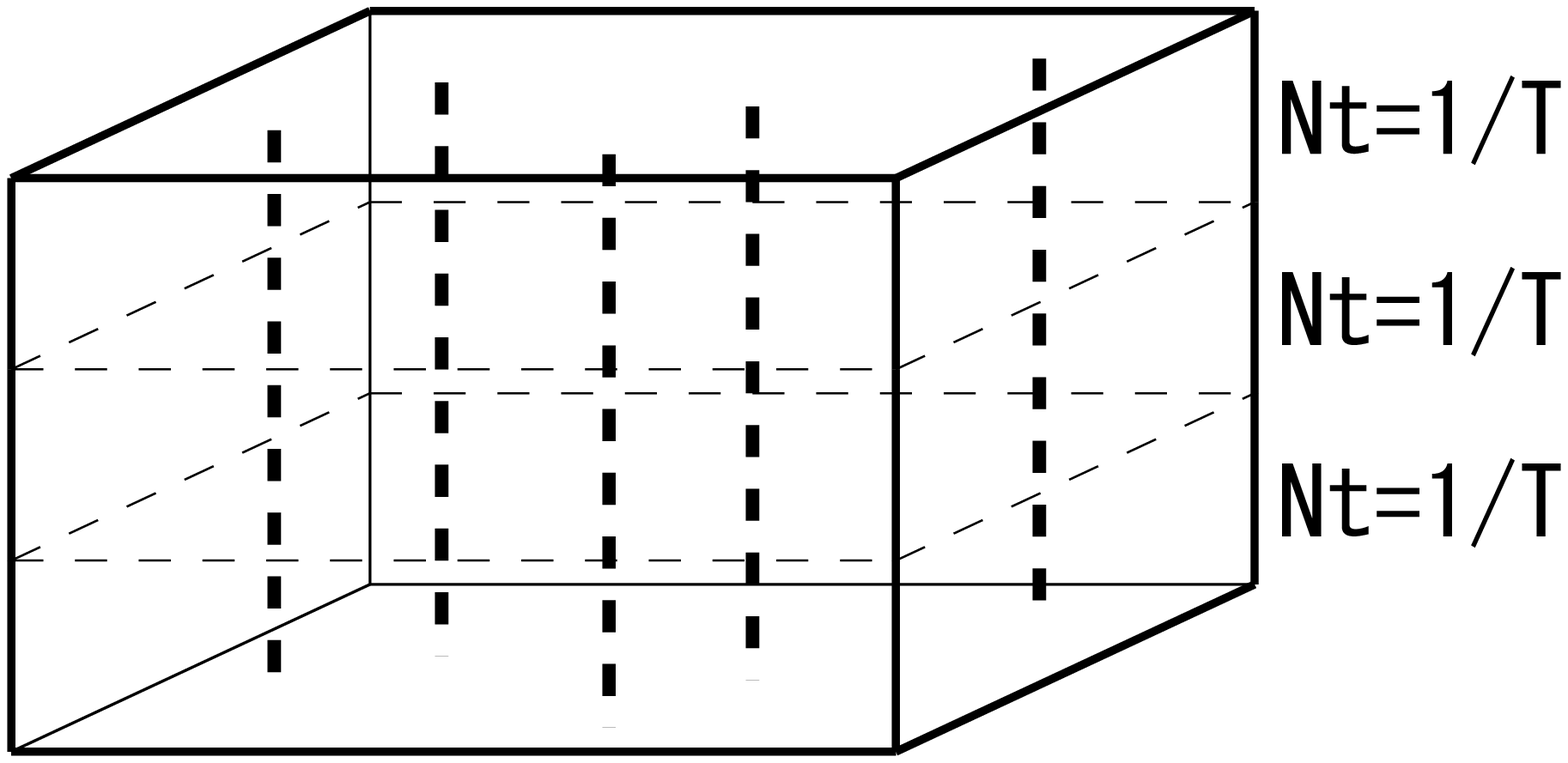}
\end{center}
\caption{\label{monopoleconf}
{\bf Left:}
Monopoles take rather static configurations with small $N_t$
(at high temperature).
{\bf Middle:}
Near-zero Dirac modes are localized avoiding monopoles.
In the presence of static monopoles,
near-zero modes would be spatially localized
and temporally delocalized.
{\bf Right:}
We artificially construct $12^3\times 12$ system
duplicating $12^3\times 4$ system.
Due to the periodicity of gauge fields,
monopole configurations on this artificial $12^3\times 12$ system
are the same as those in the original $12^3\times 4$ system.
}
\end{figure}

Monopole world lines will take
``static'' configurations at high temperature;
most of the world lines are twisted around the 4-dim torus
along the temporal direction as is
illustrated in Fig.~\ref{monopoleconf}[left].
In such a case, near-zero modes can be freely extended 
along the temporal direction, since they do not encounter monopoles
as impurities for near-zero modes, which is 
schematically illustrated in Fig.~\ref{monopoleconf}[middle].
This ``clearing up'' along the temporal direction may be the main
cause for the vanishing temporal structures of near-zero modes.
It could be also the origin of the stronger localization
properties of near-zero modes:
The clearing up leads to the irrelevance of the temporal direction
and implies that the system dimensionality
for near-zero Dirac modes is reduced approximately to three.
The near-zero modes are then expected to get strongly localized 
due to this dimensional reduction.
(Dimensionality also plays an essential role
in the localization of wavefunctions.
It is a celebrated fact that,
in a system whose dimensionality is lower than two,
any random potentials readily lead to the 
exponential localization of wavefunctions.)

\begin{figure}[h]
\begin{center}
\includegraphics[scale=0.27]{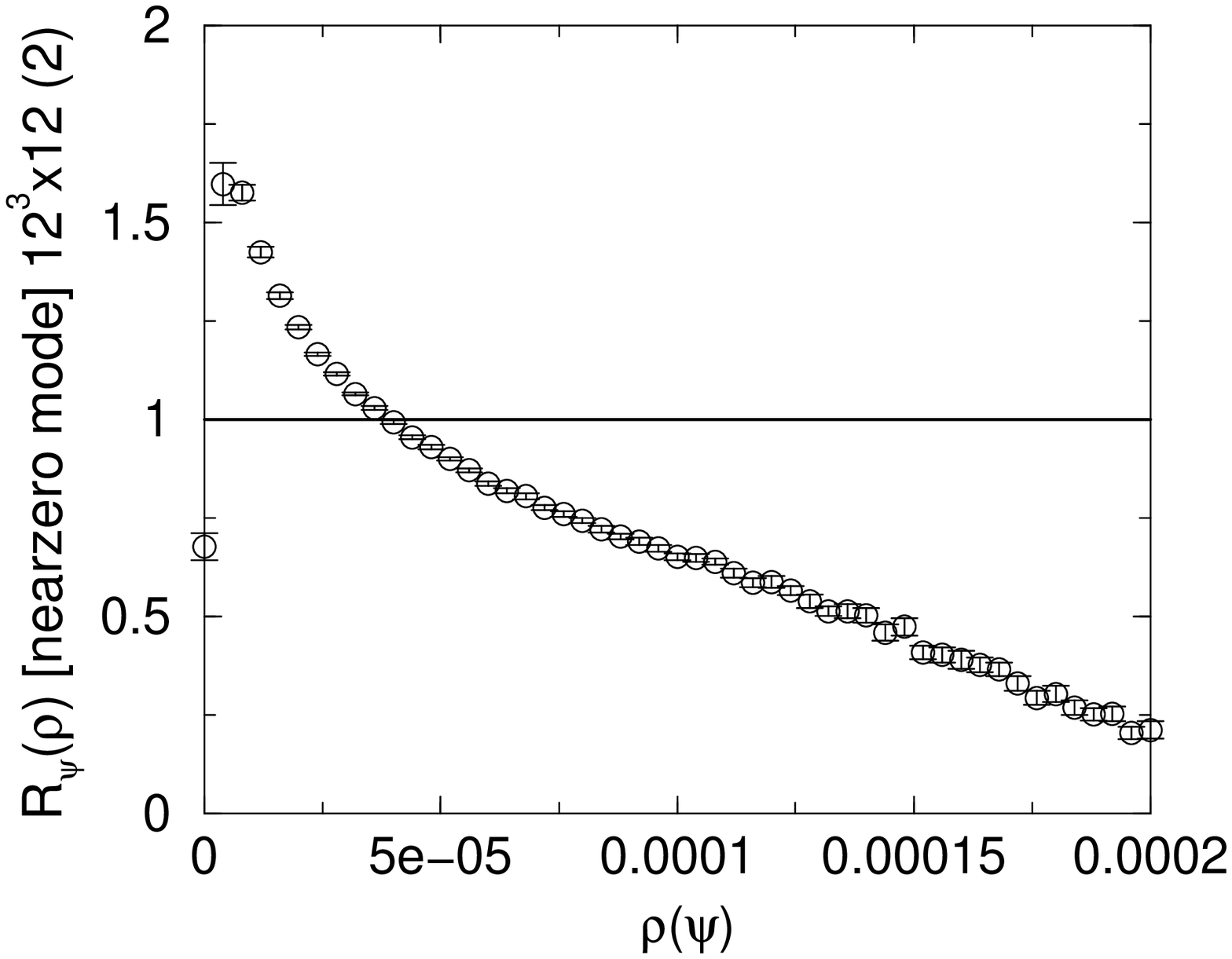}
\includegraphics[scale=0.27]{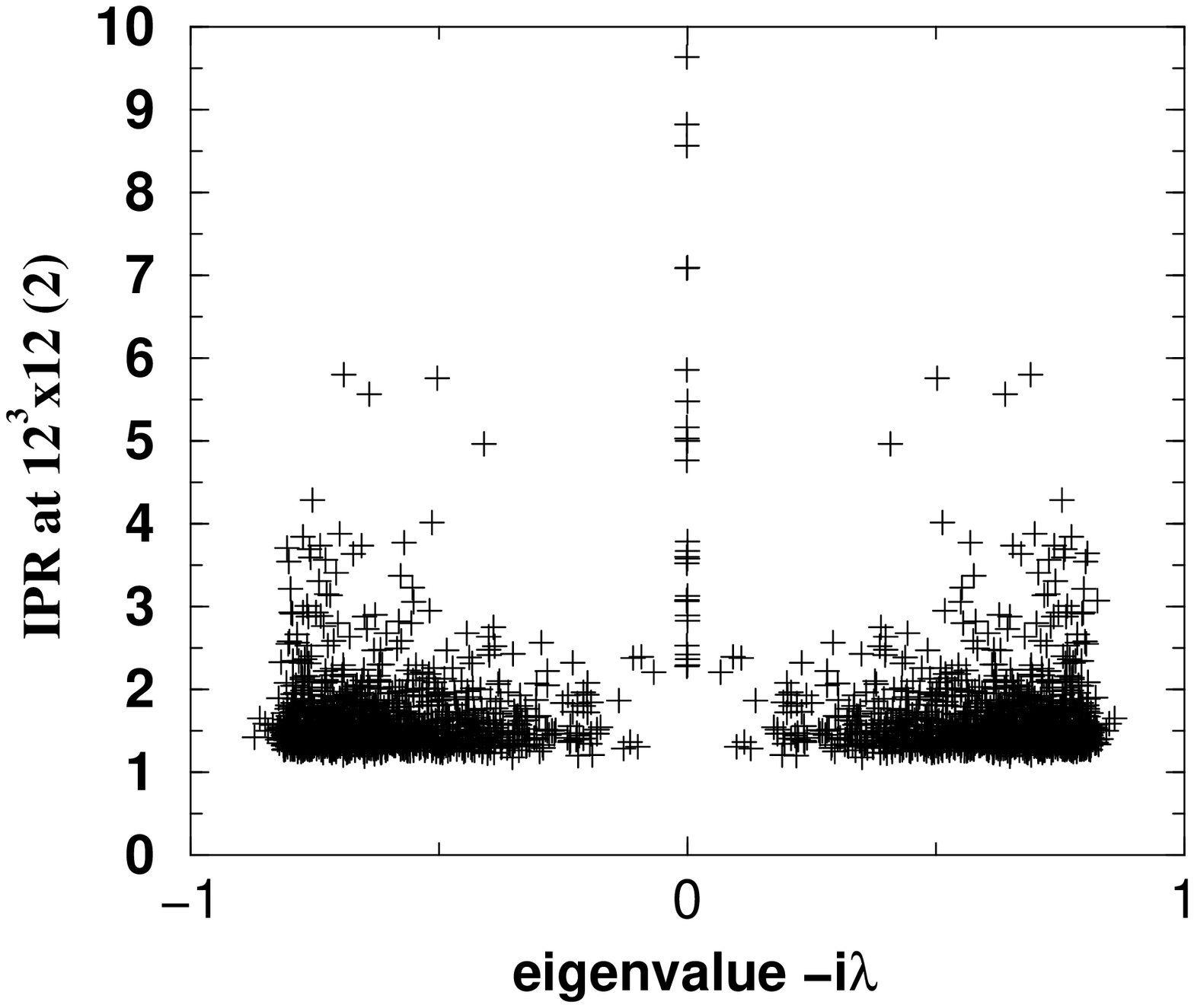}
\includegraphics[scale=0.27]{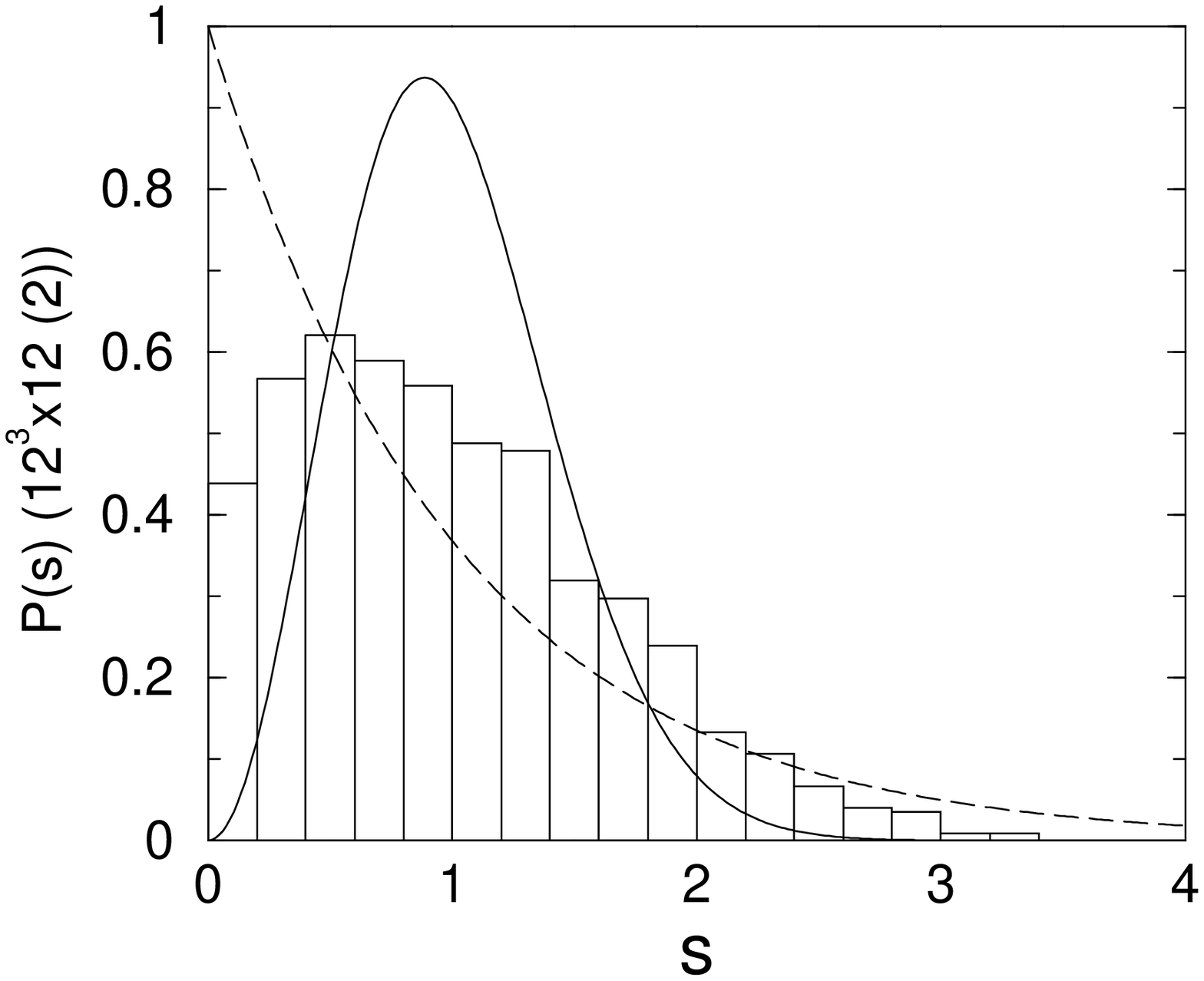}
\includegraphics[scale=0.27]{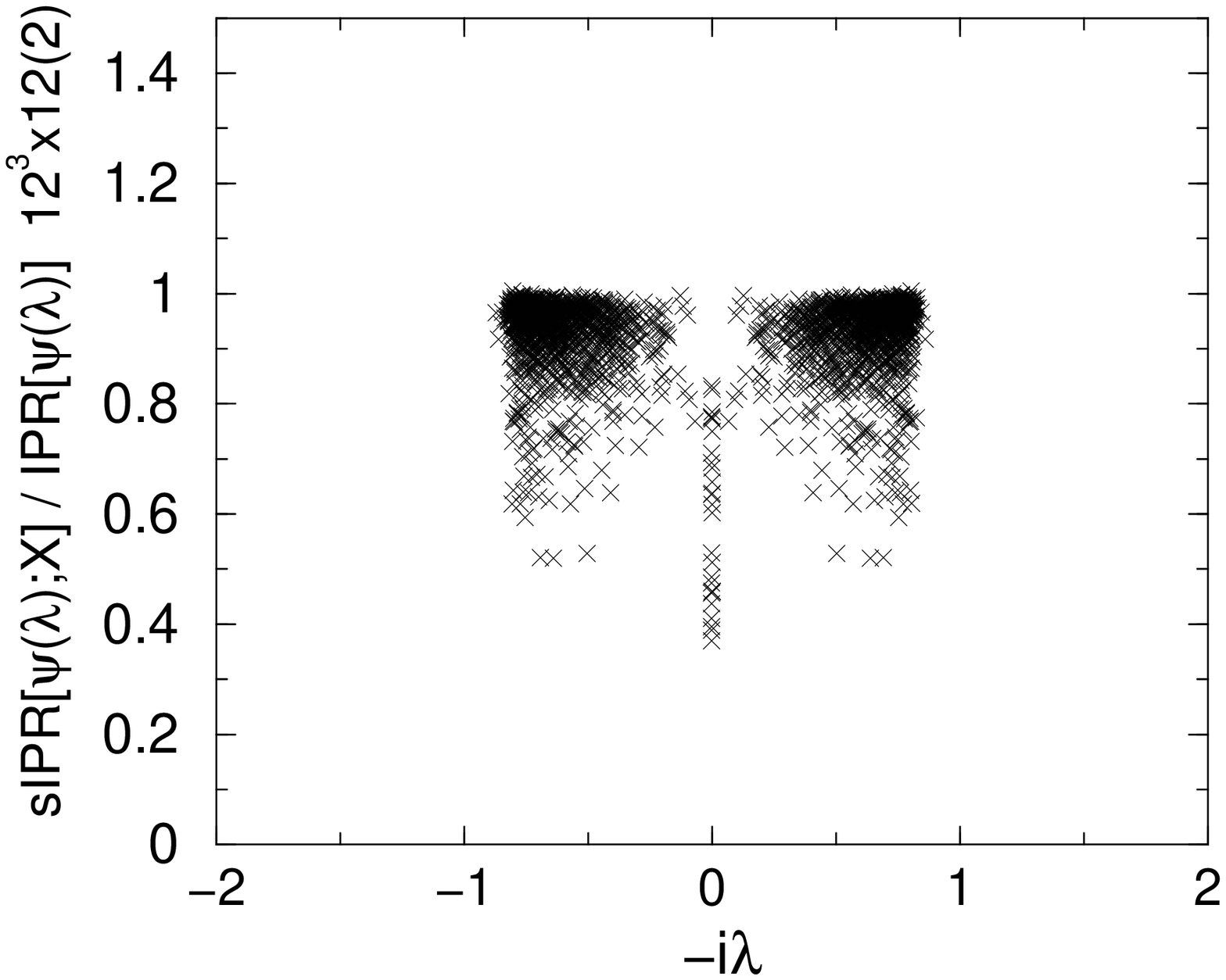}
\includegraphics[scale=0.27]{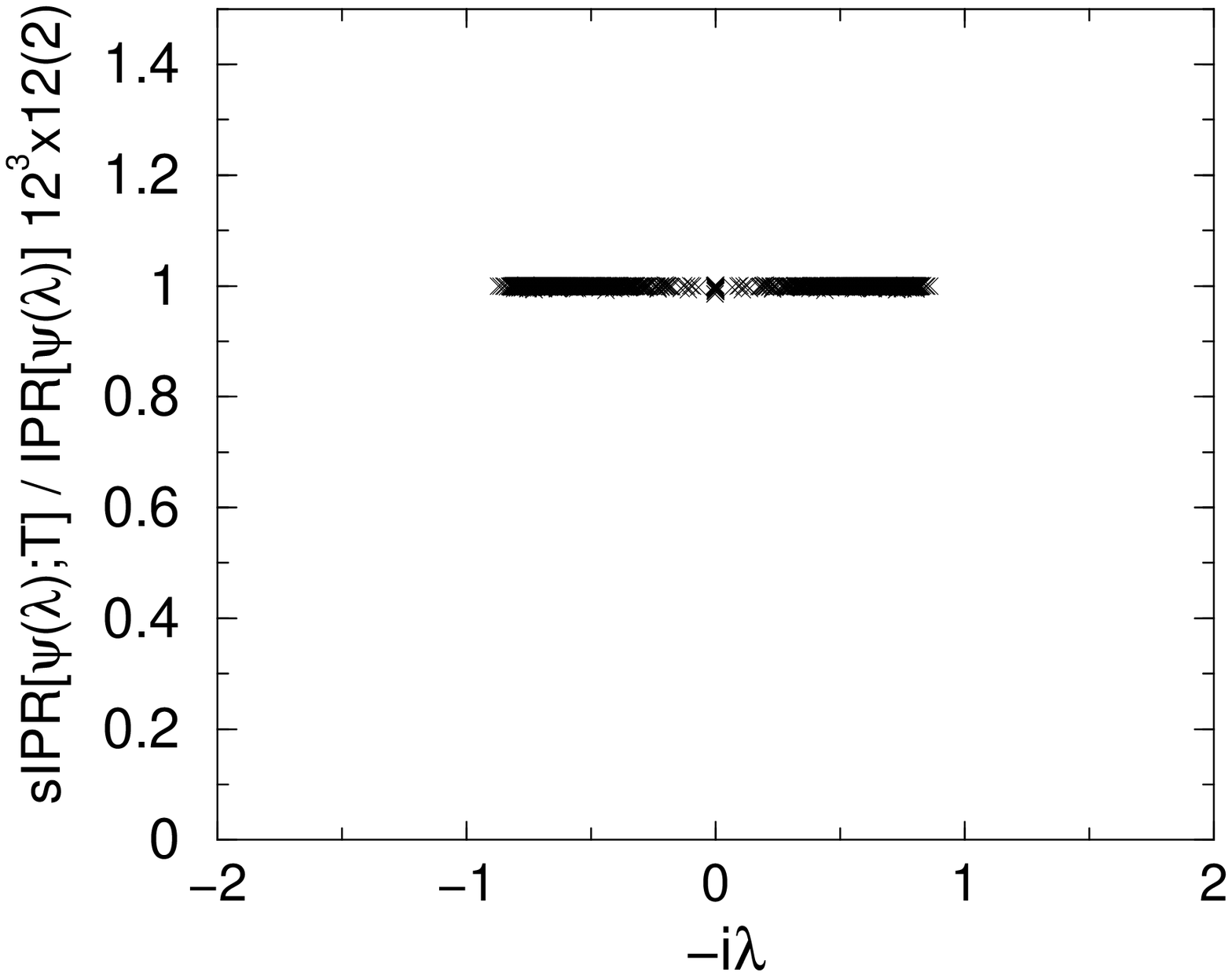}
\end{center}
\caption{\label{tpx3}
Several quantities
obtained with the artificially constructed $12^3\times 12$ lattice
are plotted;
{\bf upper left:}
The histogram ratio $R_\psi(\rho_\psi)$ for near-zero modes,
{\bf upper middle:}
the scatter plot of 
inverse participation ratios of low-lying Dirac modes,
{\bf upper right:}
the unfolded nearest-neighbor level spacing distributions $P_{\rm lat}(s)$,
{\bf lower left:}
the IPR-ratio $Is(\lambda ; i)/I(\lambda) (i= 1,2,3)$,
{\bf lower right:}
the IPR-ratio $Is(\lambda ; 4)/I(\lambda)$.
Surprisingly, all these results remain almost the same
as those in $12^3\times 4$ system.
}
\end{figure}

One may wonder which is the main origin 
of the effective dimensional reduction,
clearing up along the temporal direction
or shorter temporal extent $N_t$ (``thermal'' effect).
To answer this question,
we perform exactly the same analyses
in an artificially constructed isotropic $12^3\times 12$ system.
We define link variables $U^{(2)}_\mu({\bf x},t)$ for this new system
with $U_\mu({\bf x},t)$ in the original $12^3\times 4$ lattice.
$U^{(2)}_\mu({\bf x},t)$ is defined as
\begin{equation}
U^{(2)}_\mu({\bf x},t) \equiv
      U_\mu({\bf x},t\ {\rm mod}\ 4) \quad (0\leq t\leq 11).
\end{equation}
In other words, we construct an isotropic $12^3\times 12$ system
by piling up three $12^3\times 4$ systems.
(See Fig.~\ref{monopoleconf}[right].)
Due to the periodicity of U(1) gauge fields,
monopole configurations are the same in these two systems.
Though this newly constructed lattice is an isotropic system,
monopole world lines exhibit static configurations.
We show in Fig.~\ref{tpx3}
the histogram ratio $R_\psi(\rho_\psi)$ for near-zero modes,
the scatter plot of 
inverse participation ratios of low-lying Dirac modes,
the unfolded nearest-neighbor level spacing distributions $P_{\rm lat}(s)$,
the IPR-ratio $Is(\lambda ; i)/I(\lambda) (i= 1,2,3)$,
and the IPR-ratio $Is(\lambda ; 4)/I(\lambda)$.
Surprisingly, all the quantities essentially remain the same,
which indicates that 
the properties of low-lying modes 
do not depend on the temporal lattice extent at all,
and that the temporal dimension is actually irrelevant.

The chiral phase transition in compact QED system is controlled
by the metamorphosis of monopole world lines:
When a system is covered by large monopole clusters,
the system is complex~\cite{Takahashi:2007dv}
and the spectra of Dirac operators exhibit the Wigner distribution,
which implies strong repulsions among Dirac eigenvalues
and gives rise to the non-vanishing spectral density 
at the origin ($\langle \bar \psi \psi \rangle\ne 0$).
On the other hand, at high temperature,
monopole world lines take static configurations 
and the system undergoes the clearing up along the temporal direction,
which would lead to the effective dimensional reduction
for near-zero modes and to the exponential localization of the modes.
The exponential localization of Dirac modes causes
weaker repulsive forces among eigenvalues exhibiting
the Poisson-like distribution of Dirac spectra.
Such weak repulsive forces cannot form
non-zero spectral density at the spectral origin
($\langle \bar \psi \psi \rangle = 0$).
The metamorphosis of monopole world lines is also responsible
for the deconfinement phase transition.
In a system where large monopole clusters cover the entire volume,
it is in confinement phase.
When monopole world lines take static configurations,
the system breaks out from the confinement phase
to the deconfinement phase.
Such static monopoles still lead to the spatial confinement,
which is the remnant of the confinement
and gives rise to remaining nonperturbative phenomena.

These two phase transitions are then
considered to be both controlled by the monopole dynamics.
The coincidence of two ``transition temperatures''
could be naturally understood from this viewpoint.
At least, if the metamorphosis of monopole world lines 
in the vicinity of the phase transition point is quick enough,
the transition temperatures would be similar in magnitude.
The clarification of Polyakov loops in terms of 
Dirac spectra\cite{Gattringer:2006ci}
could also give us deeper understanding of the possible relationship
between confinement and chiral symmetry breaking.

\section{Summary}
\label{summary}

We have studied the properties of low-lying Dirac modes
in quenched compact QED at $\beta$=1.01,
employing $12^3\times N_t$ lattices 
with $N_t = 1/T = 4,6,8,10,12$.
The overlap formalism has been adopted for the fermion action.

We have found several features worth noting:
\begin{itemize}
\item
The nearest-neighbor level spacing distribution of Dirac eigenvalues
coincides with the Wigner distribution
at $N_t = 12$ (confinement phase),
and it gradually changes to the Poisson distribution as $N_t$ is decreased,
which is consistent with
the spectral density at the spectral origin.
\item
Near-zero modes exhibits stronger localization at high temperature,
which is the origin of the Poisson-like distribution of Dirac spectra.
\item
Near-zero modes lose their temporal structures above $T_c$.
\item
Near-zero modes have been found to have universal anti-correlations again
with monopole world lines below and above the critical temperature.
\end{itemize}
The chiral phase transition at ``high temperature''
in compact QED seems to be controlled
by the metamorphosis of monopole world lines.
Monopole world lines take static configurations at high-T system
and the system undergoes the clearing up along the temporal direction,
which would lead to the effective dimensional reduction
for near-zero modes.
This reduction 
is responsible for the exponential localization of the modes,
which causes
weaker repulsive forces among eigenvalues and gives rise to
the Poisson-like distribution of Dirac spectra.
Due to such weak repulsive forces,
the spectral density at the spectral origin vanishes
($\langle \bar \psi \psi \rangle = 0$).
Taking into account that
the metamorphosis of monopole world lines
is also responsible for the deconfinement transition,
we conjecture from the microscopic viewpoint
that the chiral and confinement/deconfinement transitions
in compact QED
are both induced by a single origin, monopole's dynamics.

\section*{acknowledgments}

The author thanks Dr.~K.~Fukushima for useful comments.
All the numerical calculations in this paper were carried out
on NEC SX-8 at the Yukawa Institute Computer Facility
and on NEC SX-8R at CMC and RCNP, Osaka university.
This work was supported by 
the 21st Century COE ``Center for Diversity and University in Physics'',
Kyoto University and 
Yukawa International Program for Quark-Hadron Sciences (YIPQS).

\end{document}